\documentclass[pdflatex,sn-standardnature]{sn-jnl}
\UseRawInputEncoding
\jyear{2021}%
\usepackage{multirow,bigdelim}
\usepackage{enumitem}

\usepackage[utf8]{inputenc}
\usepackage[T1]{fontenc}
\usepackage{graphicx}

\usepackage{amsmath,amsbsy}
\usepackage{multibib}
\newcites{app}{References}

\setlist{font=\normalfont\bfseries}
\theoremstyle{thmstyleone}%
\theoremstyle{thmstyletwo}%

\theoremstyle{thmstylethree}%

\newcommand{\hmpc}{\,h^{-1}\,{\rm Mpc}}
\newcommand{\hmsun}{\,h^{-1}\,M_\odot}
\newcommand{\dcirc}{^{\circ}}
\newcommand{\zobs}{\ensuremath{z^\mathrm{obs}}}
\newcommand{\zav}{\ensuremath{z=2.3}}

\begin{document}

\title[Fate of COSMOS Protoclusters]{Predicted Future Fate of COSMOS Galaxy Protoclusters over 11 Gyrs with Constrained Simulations}

\author*[1]{\fnm{Metin} \sur{Ata}}\email{metin.ata@ipmu.jp}

\author[1]{\fnm{Khee-Gan} \sur{Lee}}\email{kglee@ipmu.jp}

\author[2,3]{\fnm{Claudio} \sur{Dalla~Vecchia}}\email{caius@iac.es}
\author[2,3]{\fnm{Francisco-Shu} \sur{Kitaura}}\email{fkitaura@ull.edu.es}

\author[4]{\fnm{Olga} \sur{Cucciati}}\email{olga.cucciati@inaf.it}
\author[5,6]{\fnm{Brian C.} \sur{Lemaux}}\email{bclemaux@ucdavis.edu}
\author[7]{\fnm{Daichi} \sur{Kashino}}\email{kashino.daichi@b.mbox.nagoya-u.ac.jp}
\author[8]{\fnm{Thomas} \sur{Müller}}\email{tmueller@mpia.de}

\affil[1]{Kavli Institute for the Physics and Mathematics of the Universe (WPI), The University of Tokyo Institutes for Advanced Study, The University of Tokyo, Kashiwa, Chiba 277-8583, Japan}

\affil[2]{Instituto de Astrof\'isica de Canarias, s/n, E-38205, La  Laguna, Tenerife, Spain}
\affil[3]{Departamento  de  Astrof\'isica, Universidad de La Laguna,  E-38206, La Laguna, Tenerife, Spain}
\affil[4]{INAF - Osservatorio di Astrofisica e Scienza dello Spazio di
Bologna, via Gobetti 93/3, 40129 Bologna, Italy}

\affil[5]{Gemini Observatory, NSF’s NOIRLab, 670 N. A’ohoku Place, Hilo, Hawai’i, 96720, USA}
\affil[6]{Department of Physics and Astronomy, University of California, Davis, One Shields Ave., Davis, CA 95616, USA}
\affil[7]{Institute for Advanced Research, Nagoya University, Furocho, Chikusa-ku, Nagoya, 464-8601, Japan}
\affil[8]{Max Planck Institute for Astronomy, K\"{o}nigstuhl 17, D-69117 Heidelberg, Germany}





\maketitle

\textbf{Cosmological simulations are crucial tools in studying the Universe,
but they typically do not directly match real observed structures. Constrained cosmological simulations, on the other hand,
are designed to match the observed distribution of galaxies. Here,
we present constrained simulations based on spectroscopic surveys
at a redshift of $z \sim 2.3$, corresponding to an epoch of nearly 11
Gyrs ago. This allows us to “fast-forward” the simulation to our
present-day and study the evolution of observed cosmic structures
self-consistently. We confirm that several previously-reported pro-
to clusters will evolve into massive galaxy clusters by our present
epoch, including the ’Hyperion’ structure that we predict will
collapse into a giant filamentary supercluster spanning 100 Mega-
parsecs. We also discover previously unknown protoclusters, with
lower final masses than typically detectable by other methods, that
nearly double the number of known protoclusters within this volume. Constrained simulations, applied to future high-redshift data
sets, represents a unique opportunity for studying early structure
formation and matching galaxy properties between high and low
redshifts.
}
\small
\section*{Introduction}
Understanding the formation of large-scale structures in the Universe,  starting from tiny fluctuations in the matter density, and subsequently evolving gravitationally into the
complex cosmic web seen at the present epoch, is a key ambition of cosmological science. 
The formation and evolution of galaxy clusters throughout different cosmic epochs is a crucial probe of the current cosmological $\Lambda$CDM concordance model (see \cite{Kravtsov2012} for a review), and
is studied with a wide range of observational techniques. 

The collapse time of massive galaxy clusters is of the order of the Hubble time, thus we expect their formation to typically complete at redshifts $z\lesssim 1$. 
At higher redshifts $z \gtrsim 1.5$, we usually do not observe collapsed massive galaxy clusters but instead extended accumulations of galaxies that do not yet form bound structures \cite{Muldrew2015} or only a collapsed core of the whole structure \cite{Miller2018,Oteo2018}

These diffuse formations in the evolving cosmic web \cite{Einasto2011,Suhhonenko2011}, that correspond to Lagrangian volumes of $ V\gtrsim (10\,\hmpc)^3$ \cite{Chiang2013}, are called \textit{galaxy protoclusters} (see \cite{Overzier2016} for a review) and represent the progenitors of galaxy clusters seen in the Local Universe.
As gravitationally evolving objects, protoclusters are ideal observables to study early structure formation and to compare to theoretical predictions. Moreover, they are active
sites of star and galaxy formation during the $z\sim 2-3$ Cosmic Noon epoch \cite{Chiang2017}, which makes them excellent laboratories to jointly study the interplay between baryonic physics and dark matter models. 

Observationally, the effort to find and characterize protoclusters is a lively, ongoing field. 
In particular, the COSMOS field \cite{Capak2007} is an excellent site
for this, as it is covered by deep and coordinated multi-wavelength observations over a wide field ($> 1$ square degree, corresponding
to $>100\,$Mpc in the transverse extent) suited to protocluster studies. 

In addition, surveys using Ly$\alpha$ forest tomography --- which probes the large-scale diffuse hydrogen distribution --- have targeted parts of this field \cite{Lee2014,Lee2018,Newman2020,Horowitz2021}, offering a complementary way to study structure formation apart from galaxy surveys.
Multiple studies have identified protoclusters in COSMOS 
with various claims they are potentially progenitors of galaxy clusters such as Coma or Virgo, with present-day masses up to $\sim10^{15}\, M_\odot$  \cite{zfire1,Yuan2014,Franck2016,Diener2015,Lee2016,Chiang2015,Casey2015,Wang2016,Cucciati2018,Darvish2020,Polletta2021,Champagne2021}. However, these estimates were done with different methods that may not be consistent with each other, and typically did not take into account the large-scale structure environment over $>10 \hmpc$, which is known to determine the evolution of the protoclusters  \cite{Muldrew2015,Chiang2013,Overzier2016}.
Also, studies show that density peaks at high redshifts do not necessarily collapse into massive galaxy clusters at $z=0$. In turn, massive structures at $z=0$ do not always originate from single high redshift density peaks \cite{Cuesta2008}.
Up to this point, there has not been a uniform and self-consistent study dedicated to these structures in the COSMOS field.


We address this problem with \textit{constrained simulations} applied towards the rich legacy of large-scale spectroscopic surveys that have been conducted on the COSMOS field over nearly a decade, achieving a cosmic volume and number density unmatched anywhere else on the sky.
Standard cosmological simulations start from randomly created Gaussian matter density fluctuations 
and produce structures that are statistically consistent with the observable Universe,
but are not intended to directly match actual observed structures. 
On the other hand, constrained cosmological simulations are designed to reproduce specific large-scale structures in observations. 
However, hitherto, most applications have mainly focused on the Local Universe or nearby structures  \cite{Gottlober2010,Hess2013,Wang:2014,Jasche:2015,Libeskind2020}. 
By performing cosmological simulations constrained from high-redshift data ($z>2$), we can potentially probe epochs when cosmic structure growth
was still in the quasi-linear regime. 
This allows a more direct mapping between the initial conditions and the observed galaxy distribution than feasible
in the strongly non-linear regime of later epochs. Moreover, it opens up the exciting possibility of forward-modeling the evolution of the observed high-redshift structures to the present time, to predict whether or not they will collapse into massive cluster halos.


\section*{Constrained Simulations of the COSMOS Field}
For our analysis, we use initial density fluctuations at $z= 100$ that would evolve into a matter density field consistent with the 3D distribution of a large sample of spectroscopically-confirmed galaxies at Cosmic Noon \cite{Ata2021}. 
This sample was compiled from the zCOSMOS-Deep \cite{Lilly2006}, VUDS \cite{LeFevre2015} MOSFIRE \cite{Kriek2015}, and ZFIRE \cite{zfire1} redshift surveys over the redshift range
of $1.4\leq \zobs \leq 3.6$ and lie
within the central square degree of the COSMOS field centered on R.A.$= 150.1^\circ$, Dec$=2.2^\circ$ in the J2000 ICRS frame. 
In cosmology, the redshift is often used as a proxy for
both cosmological distance and cosmological time, which in our case is generally disconnected. To disambiguate, we use
$\zobs$ when referring to observed distance or equivalent comoving line-of-sight position in our simulations, and $z$ as the time label for the simulation. \\
Special care was taken to accurately calculate the selection functions of the individual surveys to ensure an unbiased estimate across the different surveys \cite{Ata2021}.
We then applied a multi-survey adaptation \cite{Ata2015} of the COSMIC BIRTH algorithm \cite{Kitaura2019}, which is a nested Bayesian inference algorithm for estimating the initial conditions and underlying matter density field. 
The reconstruction grid was chosen to be $256^3$ cells with a 
side length of $L_\mathrm{box} = 512\,\hmpc$, which is large enough to cover the comoving line-of-sight extent represented by 
$2.00 \leq \zobs \leq 2.52$ in COSMOS. This redshift range was specifically selected to overlap with multiple protoclusters previously reported in the literature \cite{zfire1,Yuan2014,Diener2015,Chiang2015,Wang2016,Casey2015,Lee2016,Cucciati2018,Darvish2020,Polletta2021}). 

From the posterior sample of initial conditions, we randomly selected 50 realizations to seed the simulations which were then run using the PKDGRAV3 $N$-body code  \cite{Potter2017}. 
We dub the corresponding simulation suite as ``\textbf{CO}nstrained \textbf{S}imulations of \textbf{T}he \textbf{CO}smos field" (COSTCO). 

\begin{figure}[tb]
\centering 
\includegraphics[width=.99\textwidth]{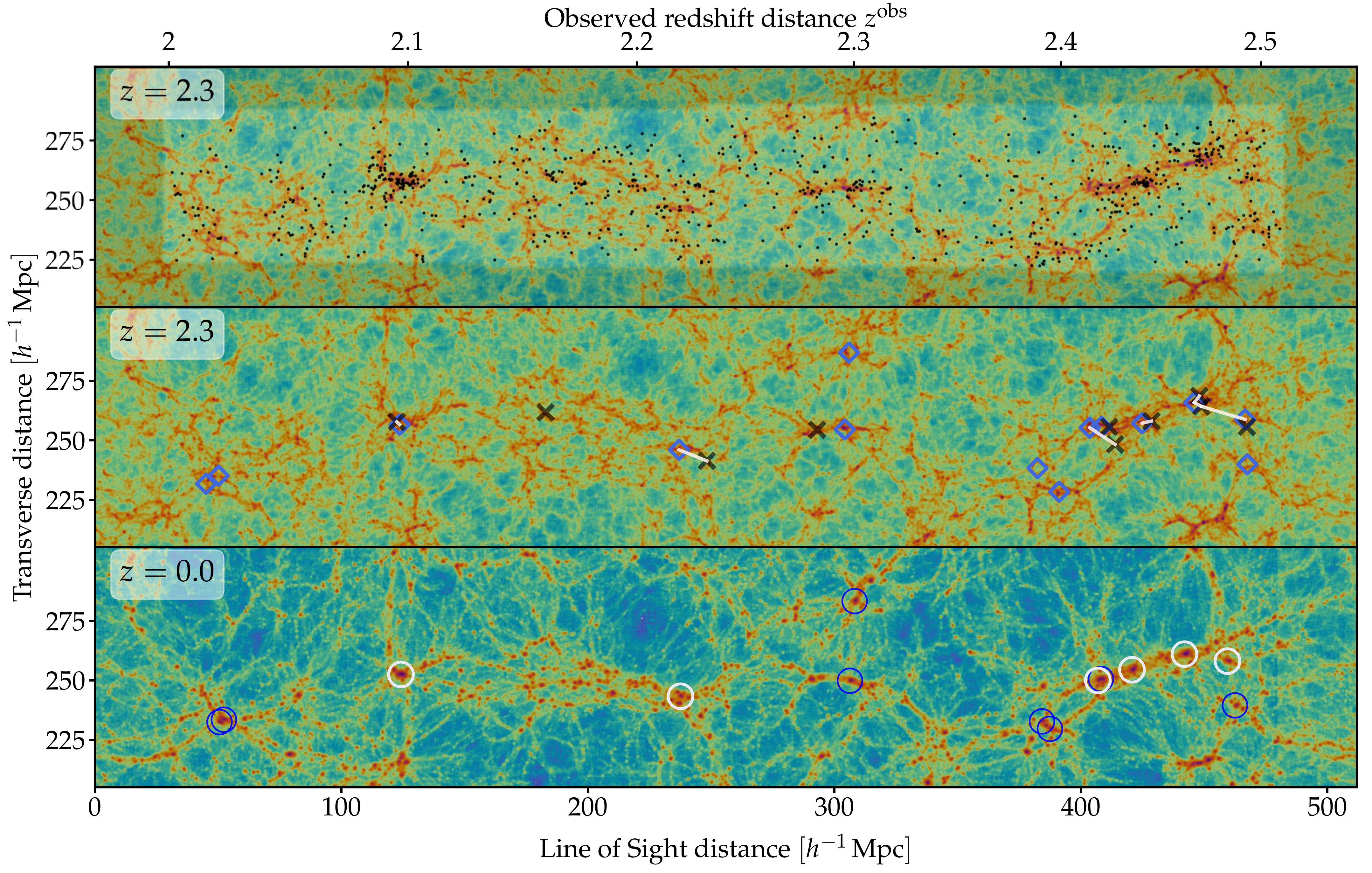}
\caption{\label{fig:dens} Matter density field of the COSMOS volume in the observed redshift range of  $2.00 \leq \zobs \leq 2.52$, from one realization out of 50 constrained simulations.  \textbf{Top \& Middle:} A $z=2.3$ snapshot of our constrained simulations. The highlighted area on the top panel indicates the observed region covered by the four galaxy surveys. Black dots show the galaxy positions. Black crosses show the \textit{Literature-Reported Protocluster
Candidates} (LRPCs), while the blue diamonds correspond to the center of mass position of protoclusters seen in COSTCO. The white lines show which LRPC is matched to each protocluster in a $r = 15\, \hmpc $ search radius. \textbf{Bottom:} The density field of the $z=0$ snapshot. The white circles indicate the halos that were successfully matched with LRPCs while blue circles show unmatched halos at $z=0$. In both slices, the ordinate axis approximately corresponds
to the Declination dimension in sky coordinates.}

\end{figure}

The COSTCO multiverse simulations were run from the initial conditions at $z=100$ until $z=0$, with 6 intermediate time snapshots being output at $z=2.0-2.5$.
Figure~\ref{fig:dens} shows projected slices of the matter density field from one of the constrained
realizations, centered at approximately R.A.=$149.99^\circ$ in the COSMOS field. We also show as fine dots the observed
galaxy positions in COSMOS, which trace the $\zav$ matter density field as expected. 

Evolved to $z=0$,
we see that the fuzzy and diffuse matter distributions at $\zav$ have
collapsed into much larger structures with higher density contrast.

\begin{figure}[tb]
\centering 
\includegraphics[width=.45\textwidth]{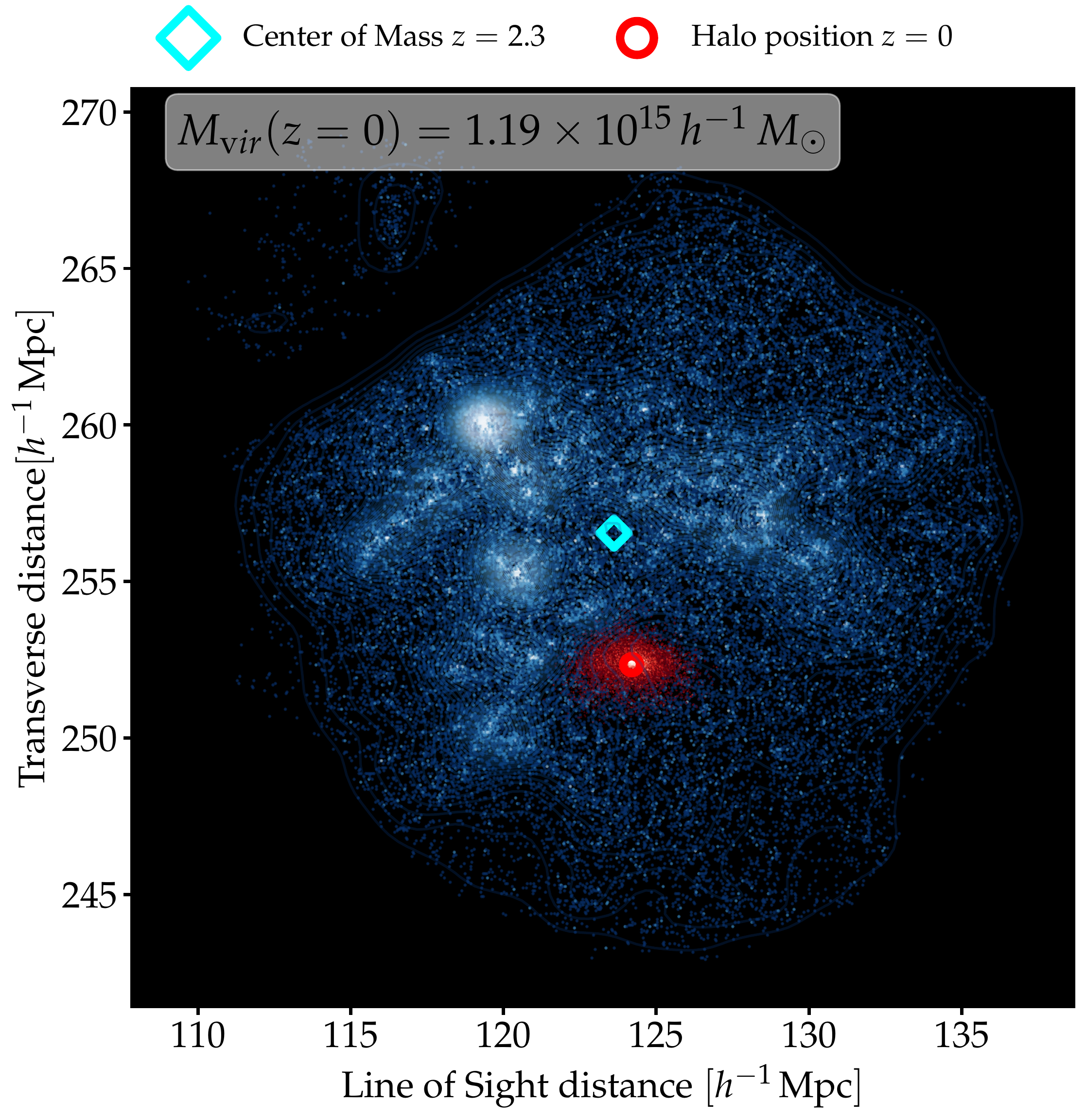}
\includegraphics[width=.45\textwidth]{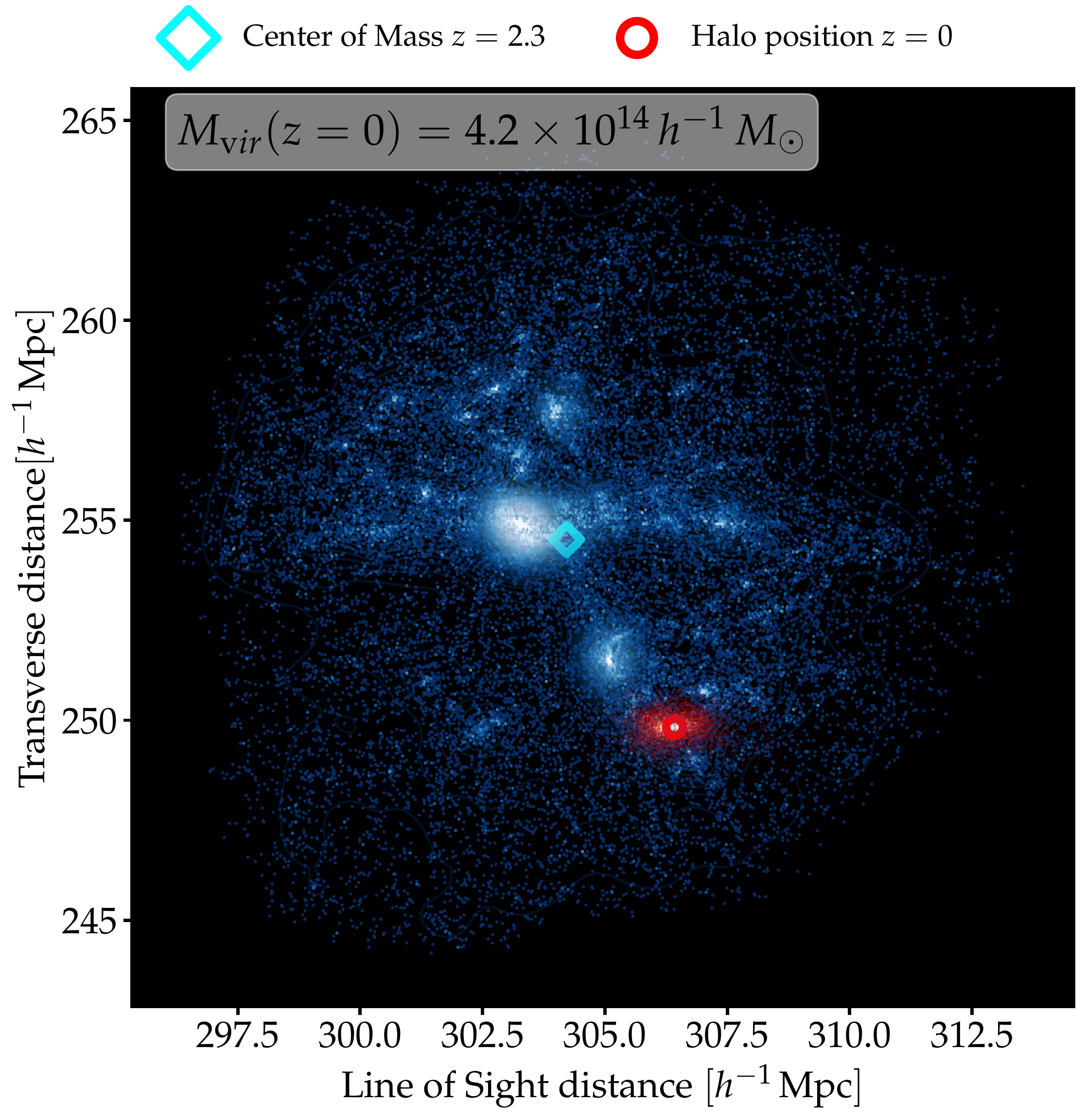}
\caption{\label{fig:halos}  Two examples of massive $M_{\rm vir} \geq 2\times 10^{14}\hmsun$ halos in our constrained simulations. The $z=0$ center of the virialized halo is shown as a red circle, with the member dark matter particles shown as red dots. The blue dots
show the same dark matter particle positions at the $z=2.3$ snapshot. Additionally, we mark the $z=2.3$ center of mass position of the dark matter particles with a cyan diamond.
The coordinate axes are the same as in the matter density plot shown in Figure~\ref{fig:dens}.}
\end{figure}

The different realizations show variation in the resulting $z=0$ filamentary structures (see Figure \ref{fig:snapshots} in the Appendix), but massive clusters form
consistently in positions corresponding to large overdensities at $\zav$.
An ensemble of simulations like COSTCO, with consistent large-scale structure and fluctuations on smaller scales is called \textit{Local Ensemble Statistics} in the literature and was first studied in \cite{Aragon-Calvo2016}.
Next, we identified collapsed halos with the ROCKSTAR halo finder \cite{rockstar}, that have a minimum virial mass of $M_{\rm vir} \geq 2 \times 10^{14}\,\hmsun$ at $z=0$, which is approximately equivalent to the mass of the Virgo cluster, the nearest massive cluster to the Milky Way.
We then traced the particles of each halo back to their Lagrangian positions at the nearest time snapshot to their observed redshift at $2.00\leq \zobs \leq 2.52$, and computed the center-of-mass position of the protocluster. 
Examples of this are shown in Figure~\ref{fig:halos}, which illustrates the extended ($\gtrsim 20\,\hmpc$) nature of galaxy protoclusters. It is interesting to note that the protocluster
center-of-mass can experience significant displacements ($\sim 10\,\hmpc$) as it evolves over 11 Gyr from $z= 2.3 $ to $z=0$. Indeed, in the case of the lower-mass cluster, the final halo position is at an Eulerian
coordinate almost outside the envelope of its $z\sim 2.3$ progenitor particles.
This illustrates the crucial role that complex gravitational interactions across the large-scale cosmic web
plays in protocluster evolution, such that oversimplified modeling with e.g. linear theory can lead to inaccurate results. The variance across different
realizations in the mature $z=0$ masses and Lagrangian center-of-mass coordinates at $\zav$ allow us to calculate uncertainties
on the position and final mass for each identified protocluster.

\section*{Results}

\begin{figure}[tb]
\centering 
\includegraphics[trim=0 4cm 1cm 2cm, clip, width=.49\textwidth]{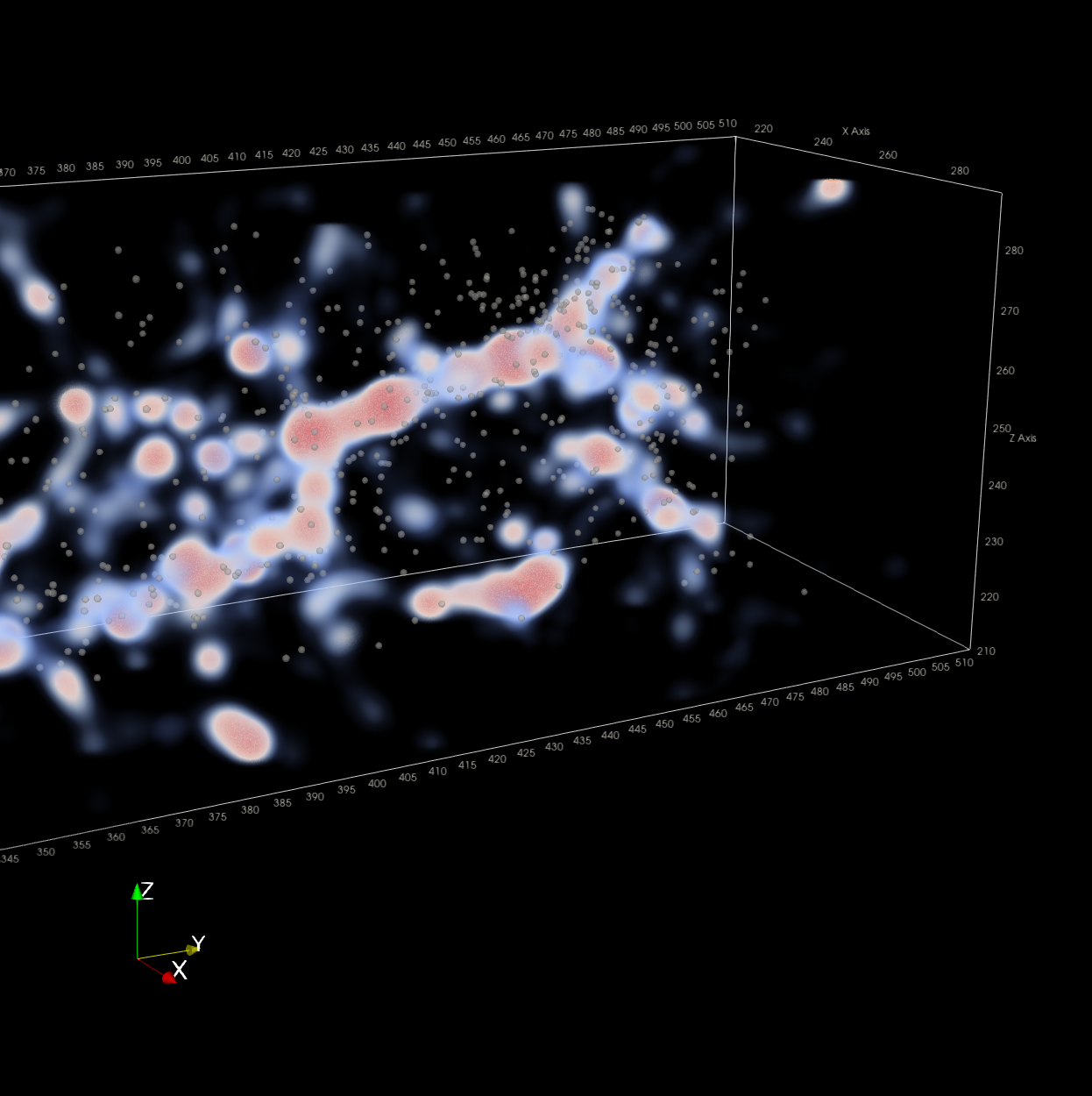}
\includegraphics[trim=0 4cm 1cm 2cm, clip, width=.49\textwidth]{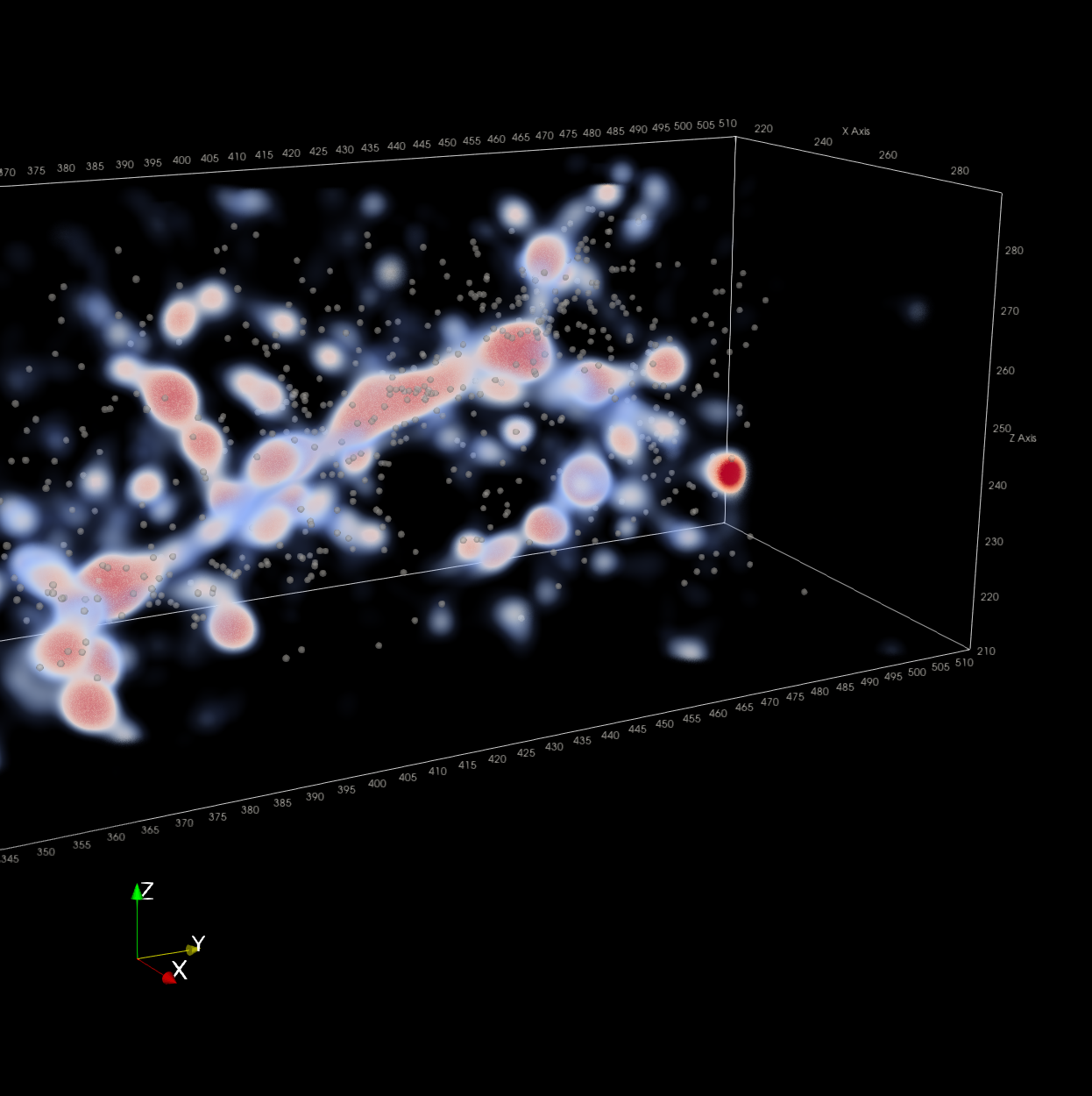}
\includegraphics[trim=0 4cm 1cm 2cm, clip, width=.49\textwidth]{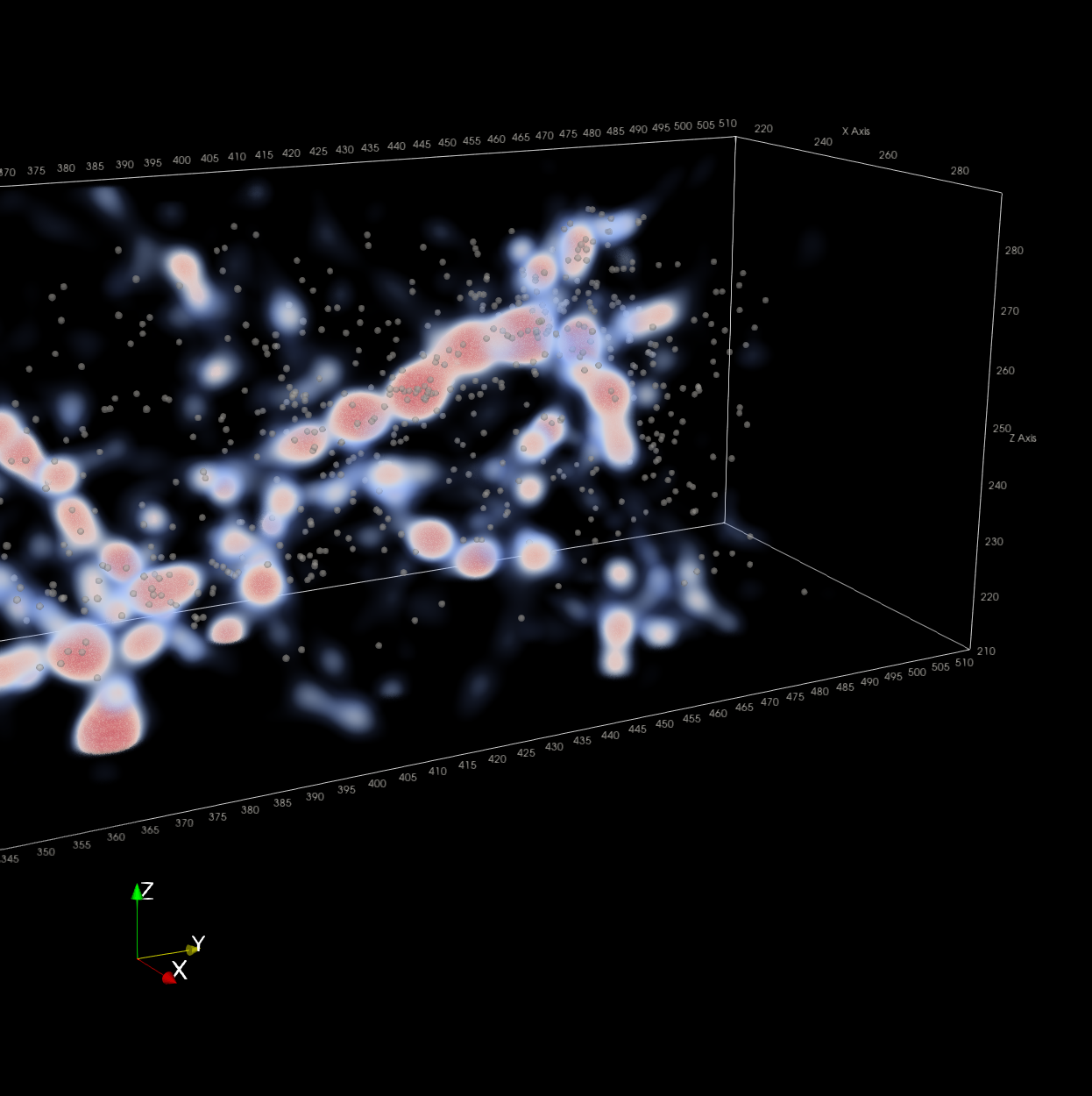}
\includegraphics[trim=0 4cm 1cm 2cm, clip, width=.49\textwidth]{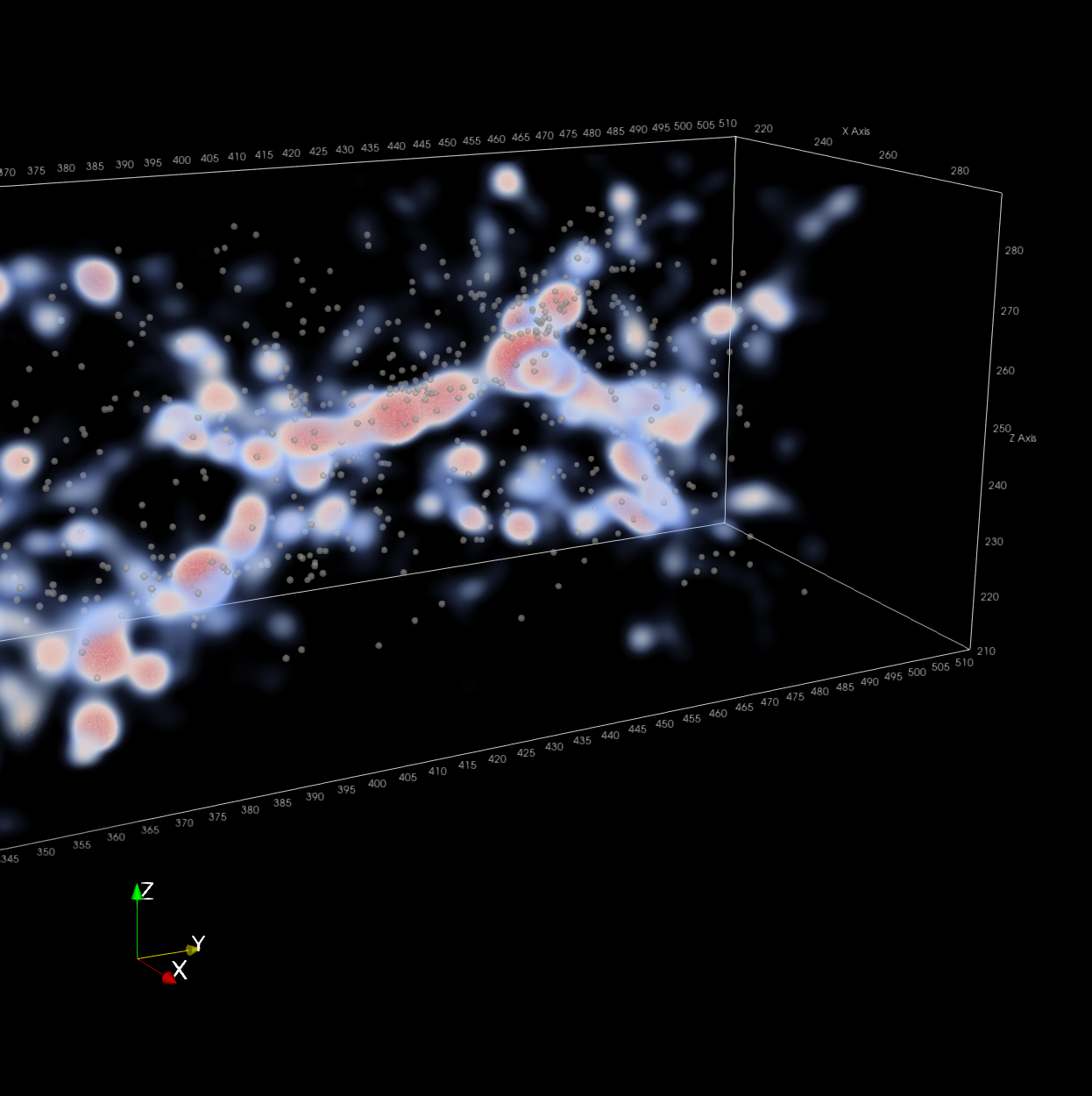}

\caption{\label{fig:hyperion} 3D visualizations of the Hyperion supercluster filament for 4 realizations at $z=0$, showing that the filamentary structure occurs consistently within our ensemble of realizations. The dots indicate
the comoving positions of the observed $\zobs \approx 2.4- 2.5$ galaxies that were used in the reconstructions. 
}
\end{figure}

To compare with observed protoclusters in the literature (which we term \textit{Literature-Reported Protocluster
Candidates} or `LRPCs'),  
we convert the reported coordinates for each LRPC into equivalent Cartesian coordinates in our simulation grid, where the observed redshift, \zobs, is converted into a line-of-sight comoving distance using a fiducial Planck 2018 cosmology \cite{Planck2018}. We then search for the center-of-mass coordinates of COSTCO protoclusters within a spherical aperture of radius $15\,\hmpc$.
The following are our constrained simulations results for the LRPCs (which are also
summarized in Table~\ref{tab:proto_obs}):

\begin{description}[leftmargin=*,labelindent=2.5em,itemsep=0.5em]
\item[ZFOURGE/ZFIRE:]
This protocluster was first identified through medium-band imaging \cite{zfire1}, and
subsequently confirmed through spectroscopic follow-up at $z=2.095$ \cite{Yuan2014}. 
We find a matched protocluster in all our realizations and predict it to evolve into a Coma-like cluster halo of $M_{\rm vir}  = ( 1.2 \pm 0.3 )\times 10^{15} \hmsun$. 
The COSTCO simulation ensemble predicts that the center-of-mass will experience only very small displacements along the line-of-sight between $z=2.1$ and $z=0,$ but there will be a significant systematic displacement toward lower declination values at $z=0$. This suggests the presence of another significant overdensity to the South of the current data sample, which we will follow-up in a separate paper.

\item[Hyperion:]
Various authors had successively discovered protoclusters at $2.423\leq \zobs \leq 2.507$,
across a region spanning approximately $\Delta \theta = 0.5\dcirc$ on the sky \cite{Diener2015,Chiang2015,Wang2016,Casey2015,Lee2016}.
It was then argued by \cite{Cucciati2018} that these protoclusters likely form a connected
structure, dubbed `Hyperion', with seven distinct density peaks that will evolve into a super-cluster by $z=0$, with a total final mass of $4.8 \times 10^{15}\, M_\odot$. COSTCO successfully identifies massive cluster halos at the reported position of Hyperion in all the realizations.
Our constrained simulations show that a partial merging of Hyperion is likely by $z=0$, but we never witnessed a complete merging of all the constituent peaks to form one cluster only. Instead,
COSTCO indicates that, on average, four virialized clusters will coalesce out of Hyperion to form a massive filamentary group of clusters with an aggregate mass of $M_{\rm vir} = (2.5 \pm 0.5) \times 10^{15}\hmsun$ within collapsed cluster halos, spanning over a distance of $d_{\rm Hyp} = (65 \pm 10) \hmpc$. The most massive halo is found to have a virial mass of $M_{\rm vir}(z=0) = (1.25 \pm 0.35) \times 10^{15}\hmsun$. The final collapsed structure is approximately equivalent in spatial extent and mass to the Coma/A1367 filament 
in the Local Universe \cite{Fontanelli:1984}.
Another analog is the elongated supercluster core of the Sloan Great Wall structure \cite{Einasto2016}, although the limited volume of the current COSMOS data does not permit us to investigate whether the analogy holds to the full scale of the Great Wall.
Examples of 3D visualizations of Hyperion are shown in Figure \ref{fig:hyperion} for 4 different realizations. 

\item[CC2.2:]
In 42 out of 50 cases, we find a cluster of $M_{\rm vir}(z=0)  =  (4.2 \pm 1.9)  \times 10^{14}\hmsun $ eventually forming out of an overdensity very close to the CC2.2 protocluster \cite{Darvish2020}. 
Our mean final halo mass for CC2.2 is less massive than estimated by \cite{Darvish2020}, although the tension is relatively weak at only $1.44\,\sigma$ and could easily be due to the different methods and data sets.

\item[G237:]
We do not find a consistently forming cluster that arises from the reported position of the recently discovered G237 protocluster, which was initially detected as a far-IR excess in the
Planck satellite data and is reported to have $M_{\rm vir}(z=0) \sim 3 \times 10^{14}\,\hmsun$. 
A closer inspection of our galaxy data shows only 2 galaxies in a sphere of 5 arcmin radius centered at R.A = 150.507$^\circ$, Dec = $2.31204\dcirc$, and $z^{{\rm obs}}= 2.16$, where G237 is observed. We analyzed the same region in our $z=2.3$ snapshots and searched for $z=0$ halos with $M_{\rm vir}\geq 2 \times 10^{14}\,\hmsun$ that had their origin in this area. Throughout the 50 realizations, this only happens twice within the observed volume.
However, we note that G237 is reported at a position that is near the margins of our observed region and covered by only the zCOSMOS-Deep survey, i.e.\ a region
where we do not expect strong constraints. 
Nevertheless, our constrained simulations only disfavor a protocluster 
at the $1.75\, \sigma$ level which is not significant especially considering the low mass reported by \cite{Polletta2021}. 

\item[CCPC-z22-006:]
At the reported coordinates by \cite{Franck2016}, namely $\rm{R.A.} = 149.93\dcirc$, $\rm{Dec} =2.2\dcirc $, and an observed redshift of $z^{\rm obs} = 2.283$, we cannot confirm a consistently forming protocluster.
In the $z =2.3$ snapshot, we can only see in 2 out of 50 cases progenitor particles
corresponding to a $z=0$ cluster within the search radius. However, in these cases the particles are associated with a nearby forming protocluster $25\,\hmpc$ away, which we will discuss below.   
Also, in contrast to G237, CCPC-z22-006 lies within the well observed region of the used galaxy surveys and therefore, our analysis tends to disfavor a protocluster at the exact position of CCPC-z22-006.

\end{description}

\begin{table}[tbp]
\centering
\begin{tabular}{l c c c l}
Name [Ref.] & R.A. [deg] &  Dec [deg]&   $z^{\rm obs}$   &     Final Mass [$h^{-1}\,M_\odot$]   \\
\hline
\hline
ZFOURGE/ZFIRE \cite{zfire1,Yuan2014} & $150.094$    &    $2.251$  &   $2.095$  & $(1.2 \pm 0.3)\times 10^{15}$        \\
\hline
 G237* \cite{Polletta2021} & $150.507$    & $2.312$  & $2.16$ &  $--$  \\ 
\hline
CC2.2* \cite{Darvish2020} &  $150.197$    &   $2.003$  &   $2.232$  & $(4.2 \pm 1.9)\times 10^{14}$\\
\hline
CCPC-z22-006 \cite{Franck2016} & $149.930$    &   $2.200$  &   $2.283$  & $--$ \\
\hline
\multirow{7}{*}{Hyperion \cite{Cucciati2018,Chiang2015,Wang2016,Casey2015,Lee2016,Diener2015}} & $150.093$    &   $ 2.404$  &   $2.468$  &  \rdelim\}{7}{20pt}[$(2.5 \pm 0.5)\times 10^{15}$]\multirow{7}{*}{ }  \\
& $149.976$    &   $ 2.112$  &   $2.426$       \\
& $149.999$    &   $ 2.253$  &   $2.444$  &                      \\
& $150.255$    &   $ 2.342$  &   $2.469$  &                         \\
& $150.229$    &   $ 2.338$  &   $2.507$  &                        \\
& $150.331$    &   $ 2.242$  &   $2.492$  &                        \\
& $149.958$    &   $ 2.218$  &   $2.423$  &         \\
\hline
\hline
 &    &    &     &  \\
\hline
\hline
COSTCO J100026.4+020940 & $150.110$    &   $2.161$  &   $2.298$  & $(4.6\pm2.2)\times 10^{14}$ \\
\hline
COSTCO J095924.0+021435  & $149.871$    &   $2.229$  &   $2.047$  & $(6.1\pm 2.5)\times 10^{14}$ \\
\hline
COSTCO J100031.0+021630 & $150.129$    &   $2.275$  &   $2.160$  & $(5.3\pm2.6)\times 10^{14}$ \\
\hline
COSTCO J095849.4+020126  & $149.706$    &   $2.024$  &   $2.391$  & $(6.6\pm2.3)\times 10^{14}$ \\
\hline
COSTCO J095945.1+020528  & $149.938$    &   $2.091$  &   $2.283$  & $(4.3\pm2.4 )\times 10^{14}$\\
\hline
\hline
\end{tabular}
\caption{\label{tab:proto_obs} COSMOS field protoclusters in the redshift range of $2 \leq z^{\rm obs} \leq 2.52$.  
For protoclusters previously reported in literature, we report the published coordinates but the final mass is from our analysis. An asterisk (*) denotes that our study does not include the recent spectroscopic follow-up observational data used for protoclusters G237 and CC2.2. The dashed line ($--$) denotes that our analysis does not find a $z=0$ halo consistently forming at the reported position. The lower part of the table summarizes the protoclusters found in this work with the $z=2.3$ center-of-mass positions and the estimated $z=0$ final masses.   }
\end{table}

\begin{figure}[tb!]
\centering 
 \includegraphics[width=1.0\textwidth]{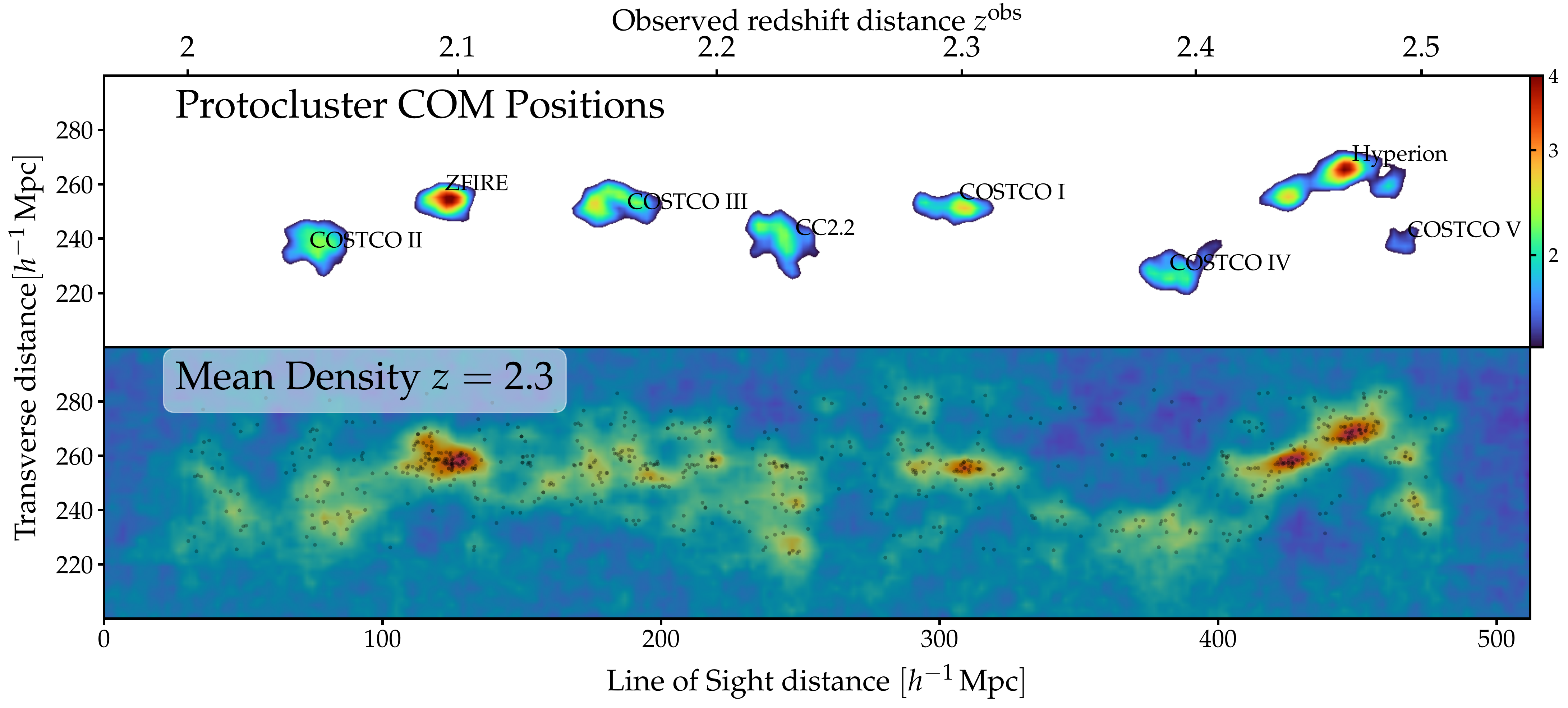}
\caption{\label{fig:heatmap}
\textbf{Top:} Confidence contours of the identified protoclusters (center-of-mass positions) with the DBSCAN method for all 50 realizations stacked into the same volume. ZFIRE and Hyperion show clearly the strongest recurrence, while the contours for COSTCO protoclusters are less distinct. \textbf{Bottom:} Averaged density field of all 50 COSTCO $z=2.3$ snapshots. The galaxy positions are superimposed with back dots. The mean density and protocluster contours agree well with the galaxy distribution at $z=2.3$.}
\end{figure}

Apart from the LRPCs, we also find several protoclusters in the COSMOS field that are, to our 
knowledge, newly-discovered through COSTCO. These typically
do not show up as strong overdensities at $2.00\leq\zobs\leq2.52$, but rather as extended milder overdensities that nevertheless contain sufficient enclosed mass to collapse into clusters by $z=0$. 
This is a new class of extended lower-mass protoclusters  --- with Virgo-like final masses of  $<10^{15}\,M_\odot$ --- that have previously eluded detections but is now revealed by our
combination of large-scale spectroscopic data and constrained simulation modelling. Since the observational data has slightly less constraining power 
at these milder overdensities, this leads to larger relative errors
in the derived masses. Nevertheless, all the newly-discovered protoclusters are robust $\sim 5\sigma$ detections
that show up in the majority of the constrained realizations: 

\begin{description}[leftmargin=*,labelindent=2.5em,itemsep=0.5em]

\item[COSTCO J100026.4+020940:]
In 48 out of 50 cases, we find a cluster progenitor forming at $\rm{R.A.} = 150.110\dcirc \pm 0.042\dcirc$,  $\rm{Dec} =   2.161\dcirc \pm 0.040\dcirc$, and  $z^{\rm obs} =  2.298 \pm  0.007$, with a final mass of $ M_{\rm vir}(z=0)   = (4.6 \pm 2.2)  \times 10^{14} \hmsun $. 
An overdensity of galaxies was first noted at this location by \cite{Lee2016}, but they did not pursue a
more detailed analysis.
This halo forms $\sim 28 \hmpc$ away from the reported position of CCPC-z22-006. In the $z=2.3$ snapshot, we noticed that the progenitor particles corresponding to COSTCO J100014.4+020748 and CCPC-z22-006 were partially overlapping in 2 out of these 48 cases. However, the resulting cluster emerged at the position of COSTCO J100014.4+020748.

\item[COSTCO J095924.0+021435:]
In 42/50 realizations we find a protocluster at $\rm{R.A.} = 149.871\dcirc \pm  0.055\dcirc$, 
$\rm{Dec} = 2.229\dcirc \pm  0.069\dcirc $, at a distance of  $z^{\rm obs} =   2.047 \pm  0.008$. The final mass is $  M_{\rm vir}(z=0) = (6.1 \pm 2.5) \times 10^{14} \hmsun$.


\item[COSTCO J100031.0+021630:]
In 49/50 cases we see a cluster progenitor forming at $\rm{R.A.} =  150.129\dcirc \pm 0.051\dcirc $, $\rm{Dec} =    2.275\dcirc \pm  0.082\dcirc$, and  $z^{\rm obs} =2.160 \pm  0.009$. We find a $z=0$ mass of $M_{\rm vir} = (5.3 \pm 2.6) \times 10^{14} \hmsun$.


\item[COSTCO J095849.4+020126:]
We find a consistently forming protocluster at $\rm{R.A.} = 149.706\dcirc \pm  0.053\dcirc  $,   $\rm{Dec} =   2.024\dcirc \pm  0.081\dcirc $, and observed redshift of  $z^{\rm obs} = 2.391 \pm  0.008 $ in 46 out of 50 simulations, with a final mass of $M_{\rm vir} (z=0)=  (6.6 \pm 2.3)  \times 10^{14} \hmsun $.


\item[COSTCO J095945.1+020528:]
We find a protocluster slightly to the South of Hyperion, which forms as a separate cluster in 27 out of 50 cases at
$\rm{R.A.} =  149.938\dcirc \pm  0.105\dcirc$,
$\rm{Dec} =   2.091\dcirc \pm  0.081\dcirc $, and
$z^{\rm obs} = 2.495 \pm  0.006 $. Our simulations predict an average mass of
$M_{\rm vir}(z=0)  = (4.3 \pm 2.4)\times 10^{14} \hmsun$. Closer inspection  reveals that even though in 40 realizations at $\zobs=2.48$ a clear overdensity forms in this vicinity, a cluster emerges only in 27 cases whereas in the other cases the overdensity is tidally disrupted by the gravitational
drag of Hyperion. 
This cluster can potentially become a substructure of the Hyperion supercluster.

\end{description}
We show the corresponding confidence map of the identified
protoclusters in the top panel of Figure \ref{fig:heatmap} , numbered in the same order
as in Table \ref{tab:proto_obs}. The bottom panel of Figure \ref{fig:heatmap} shows the mean density of
all 50 COSTCO $z=2.3$ snapshots with the galaxy positions overplotted as black dots.

\section*{Summary \& Discussion}
Based only on analysis of existing large-volume spectroscopic surveys, we have increased the number of known galaxy protoclusters at $2< z < 2.5$ in the COSMOS field
by 1.5-2$\times$, while pushing down to final masses of $M_{\rm vir}(z=0) \approx 4 \times 10^{14} \hmsun$. 
This represents a lower mass detection limit than hitherto feasible with  survey data sets, from which protocluster detections were typically in the range of $M_{\rm vir}(z=0) \sim 10^{15} \hmsun$ \cite{Overzier2016}. 
This ability to detect and simultaneously model more representative lower-mass protoclusters with large-volume surveys will allow a more uniform characterization of their demographics. 
In addition to the new objects, we have confirmed and directly modeled the fate of several previously-known galaxy protoclusters in the volume. The Hyperion aggregation \cite{Cucciati2018}, in particular, had been the subject of scientific and public curiosity: its elongated, extended nature made it difficult to ascertain whether it would collapse into a singular massive cluster by $z=0$. Our results clearly show that it will not, and instead points to a fate as multiple massive cluster cores embedded within a giant filamentary supercluster.

These results illustrate the efficacy of constrained simulations applied to high-redshift data, although it does require wide and deep spectroscopic surveys that do not currently exist elsewhere on the sky apart from COSMOS. However, in the next few years, wide-field massively multiplexed fiber spectrographs deployed on 8m-class telescopes, such as the Subaru PFS \cite{sugai:2015} and VLT-MOONS \cite{moons:2012}, will carry out large-scale high-redshift surveys across angular footprints $\gtrsim 10\times$ larger than the combined COSMOS data set that we used in our analysis. 
These upcoming surveys will allow constrained realization techniques to be even more effective in connecting cluster formation across cosmic time as well as the evolution of 
their constituent galaxies and gas. Constrained simulations of these upcoming high-redshift galaxy surveys will also allow us to probe early structure formation for consistency with the $\Lambda$CDM model with increasing sensitivity to lower mass galaxy (proto)clusters. 
Moreover, the identification of individual protoclusters (as presented in this work) combined with complementary studies such as X-ray and Ly$\alpha$ observations will allow us to study gas accretion of star-forming galaxies within these protoclusters and can provide vital model constraints for high precision hydrodynamical simulations \cite{Overzier2016}. Each identified protocluster represents a unique environment to study the morphologies and merger rates of member galaxies in high-density environments at $z\geq2$, which is, to date, only possible in theoretical studies or random cosmological simulations.

\bibliography{sn-bibliography}


\renewcommand{\figurename}{Supplementary Figure} 

\section*{Methods}

\subsection*{Input Galaxy Surveys}
\label{sec:surveys}

In this section, we summarize the galaxy catalogs that have been used in this study to compute the initial conditions. 
In total, we employed the data of four distinct spectroscopic galaxy redshift surveys, namely:
\begin{itemize}
	\item The new release of the zCOSMOS-Deep \citeapp{Lilly2006} catalog, in which the astrometry has been updated to match the COSMOS2015 photometric galaxy catalog \citeapp{Laigle2016}. It also incorporates detailed visual inspections of the spectra that has led to more reliable redshift confidence flags compared to the original catalog (Lilly et al., 2021 in prep).
	\item  The VIMOS Ultra Deep Survey (VUDS) \citeapp{LeFevre2015} catalog, which was also covered substantial parts of the COSMOS field. The galaxies were originally selected for spectroscopy based on photometric redshifts from \citeapp{Ilbert2013}.
	\item  The final data release from MOSFIRE Deep Evolution Field (MOSDEF) Survey \citeapp{Kriek2015}, whose target selection is based on Hubble Space Telescope grism observations \citeapp{Brammer2012}.
	\item The KECK/MOSFIRE Spectroscopic Survey of Galaxies in Rich Environments at $z\sim 2$ (ZFIRE) \citeapp{zfire2}, which is a spectroscopic follow-up of the ZFOURGE near-IR medium-band imaging program \citeapp{zfourge}.
\end{itemize}

\begin{figure}[tbp]
\centering 
\includegraphics[width=.49\textwidth]{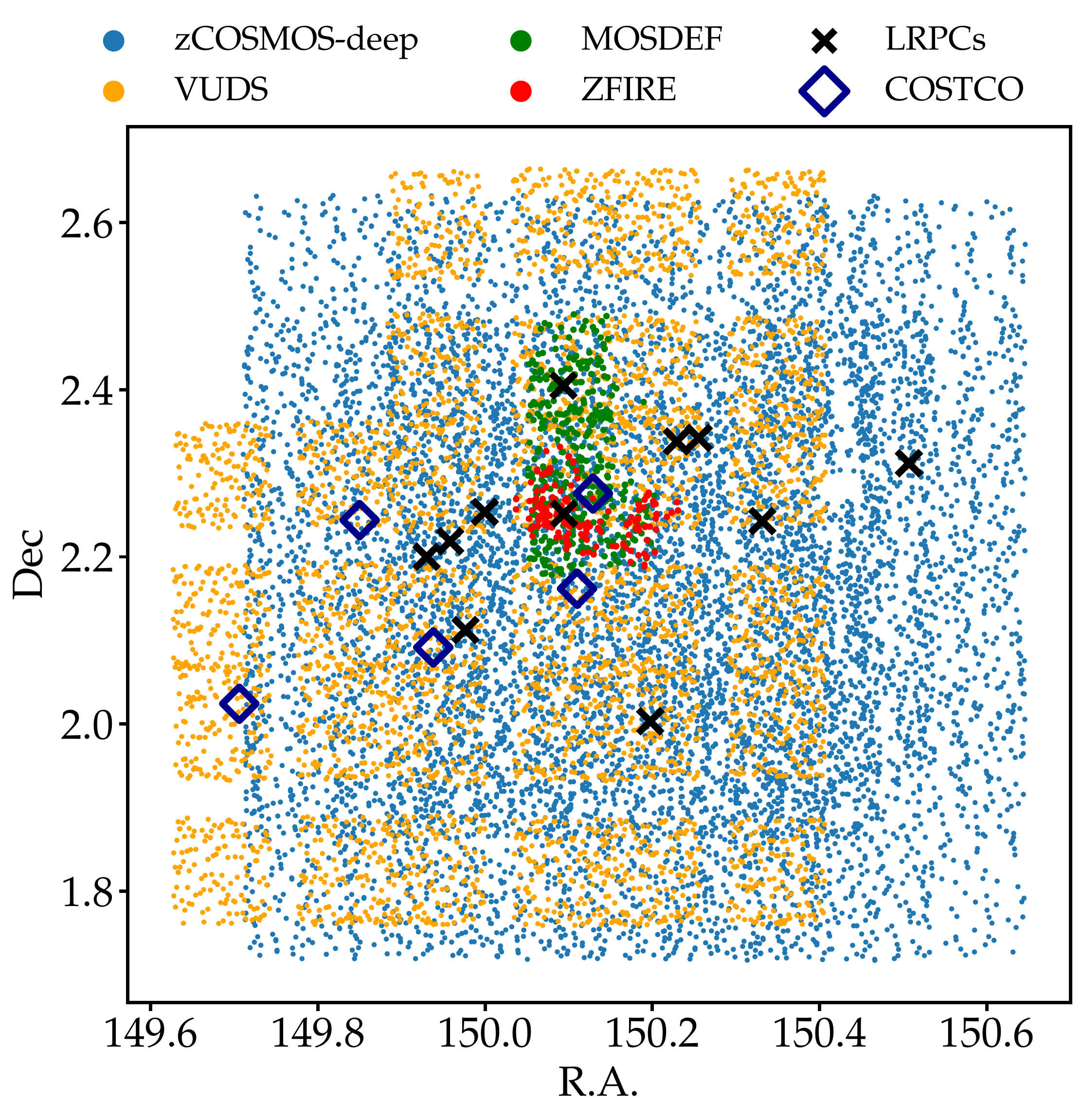}
\hfill
\includegraphics[width=.49\textwidth]{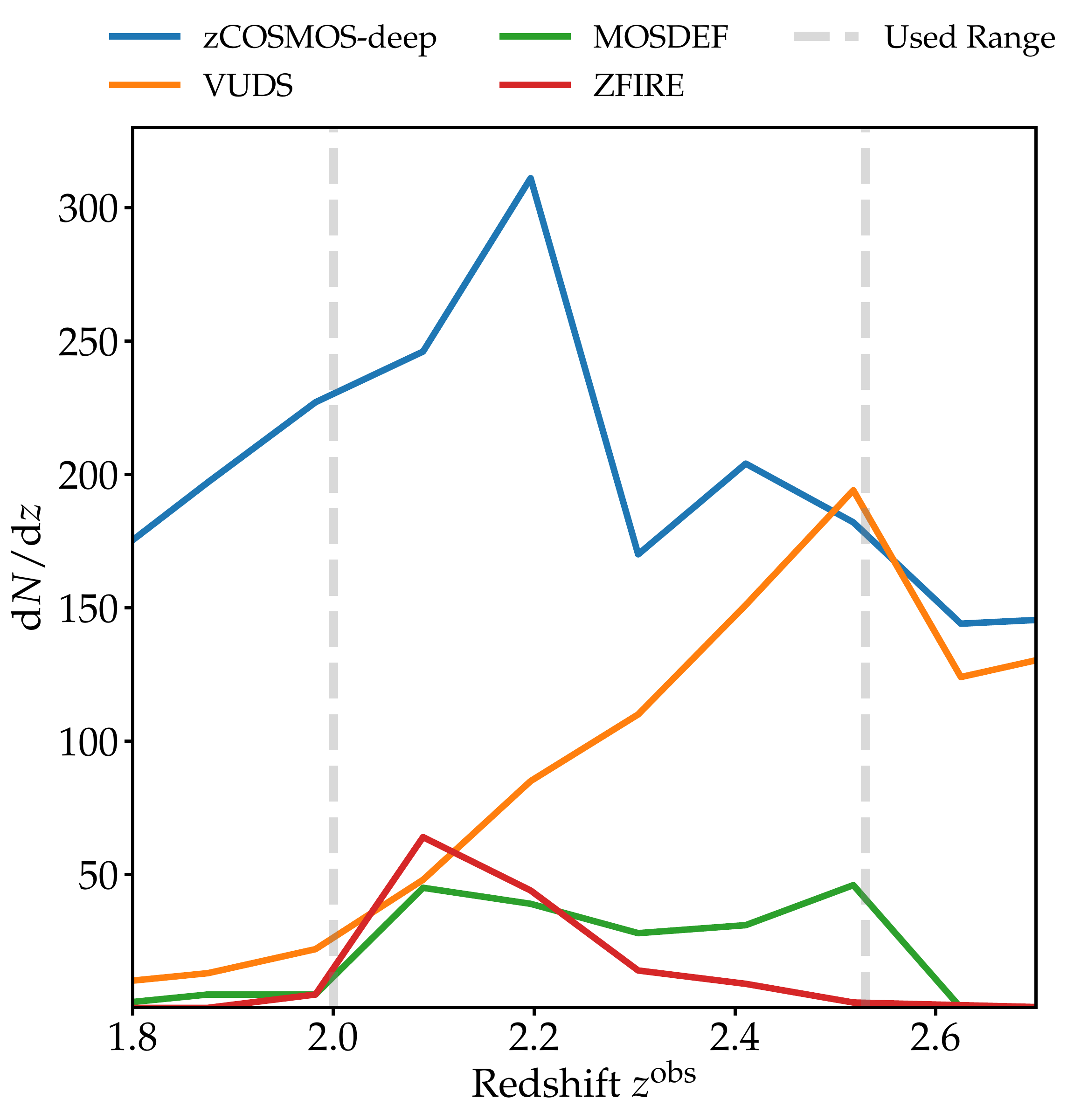}
\caption{\label{fig:surveys} \textbf{Left}: Galaxy positions on the sky for the four different surveys. The black crosses show the observationally reported overdensities.
The blue diamonds are protoclusters identified in this work, dubbed as COSTCO. \textbf{Right}: Redshift distribution of the four surveys in the same colour code as the left plot. We chose a bin width of $\Delta z^{\rm obs} = 0.1$. The grey dashed lines indicate the used redshift range.}
\end{figure}

Among the different surveys, we consider galaxies that are closer than $ \theta < 0.1^{\prime\prime}$ on the sky and additionally have a redshift separation of less than $\Delta z^{\rm obs} < 0.01$ as duplicates and remove the one with lower redshift confidence. 
The corresponding footprints and galaxy number density as a function of redshift are shown in Supplementary Figure \ref{fig:surveys}. Further, we show the angular positions of the Literature Reported Protocluster Candidates (LRPCs) and the additional COSTCO protocluster positions (this work) as black crosses and blue diamonds, respectively. 
The redshift selection criteria and the calculations of the survey selection functions are described in detail in \citeapp{Ata2021}.

\subsection*{Initial Conditions}
\label{sec:ICs}

In this section, we describe the initial conditions that were used to seed the constrained simulations.

In contrast to standard simulations starting from randomly-generated Gaussian fluctuations, \citeapp{Crocce2006, Hahn2011, Pilipenko2017, Tatekawa2020}, we seek to find initial conditions that resemble the COSMOS galaxy distribution.

First attempts to infer initial conditions from Gaussian random fields came into being nearly three decades ago  \citeapp{Dekel1990,Hoffman1991,Gramann1993,Kolatt1997} as starting point for cosmological simulations. Even though conceptually pioneering works, these methods suffered from low resolution and over-simplified assumptions, impeding further scientific applications. The technique has received renewed interest over the past decade thanks to advances in computing hardware as well as novel sampling techniques \citeapp{Jasche2013,Wang:2014,Ata2015,Jasche:2015}.

In \citeapp{Ata2021} the relevant initial Gaussian density field $\delta(\boldsymbol q)$ for this work was reconstructed corresponding to the COSMOS field using a modified multi-survey version of the COSMIC BIRTH code \citeapp{Kitaura2019}, covering the observed redshift range $1.4\leq z^{\rm obs}\leq3.6$. COSMIC BIRTH relies on the Hamiltonian Monte-Carlo (HMC) sampling \citeapp{Neal2012} technique, working on a discretized volume where each voxel of the reconstructed density field is a parameter of the posterior probability function. 
The algorithm estimates the posterior distribution of initial conditions that, after gravitationally evolving, would resemble the binned galaxy density field. 
The binned galaxy density field is therefore connected to the initial density field via a bias model, and the survey selection function at each grid cell.
This process is fully
described in \citeapp{Ata2021}. We computed five separate HMC chains with COSMIC BIRTH, starting at different initial seeds. Each chain ran at the minimum for $\sim10,000$ iterations.
After the burn-in phase, when the Markov Chain Monte Carlo (MCMC) reaches the stationary distribution, we randomly selected 50 individual realizations. We verified that the iterations were not closer than 500 MCMC steps within the chain to guarantee independent samples and accurately evaluate the posterior distribution.

We perform the initial conditions inference in a cubical box with $L_{\rm{Box}}=512\,\hmpc$ comoving side length and a mesh grid of $N_{\rm C} = 256^3$, resulting a  cell resolution of $d_{\rm{L}}=2\hmpc$.
For the current analysis, we follow the same overall method as in \citeapp{Ata2021} but restrict the reconstruction region to a redshift range of $2.00\leq z^{\rm obs} \leq 2.52$, in which a large number of reported protoclusters are located.
We transform the coordinate system of the surveys so that they are aligned with the Cartesian axes of the COSMIC BIRTH box in the plane-parallel approximation and place the data region in the center of our reconstructed volume.
The initial density fields are then calculated corresponding to a redshift of $z=100$.

To translate the initial density fields $\delta(\boldsymbol q)$ into initial seeds for cosmological simulations, we compute the three-dimensional white noise field $wn(\boldsymbol{k})$ in Fourier space, defined as:

\begin{equation}
    wn(\boldsymbol{k}) = \frac{\delta(\boldsymbol{k})}{\sqrt{P(\boldsymbol{k})}} \, ,
\end{equation}
where $\delta(\boldsymbol{k})$ is the Fourier-transformed initial density field and $P( \boldsymbol{k} )$ is the theoretical three-dimensional $\Lambda$CDM power spectrum obtained with the CAMB code \citeapp{Lewis2000}.
We read the white noise field into the MUSIC code (MUlti-Scale Initial Conditions) \citeapp{Hahn2011}.
MUSIC places a dark matter particle in each cell of the initial density mesh grid and uses the density contrast to calculate the initial displacement and velocity field for each particle with second order Lagrangian perturbation theory. 

\subsection*{Constrained Simulation Methodology}
\label{sec:sims}

Over the last decade constrained simulations have helped to understand structure formation and the dynamics of the local Universe \citeapp{Sawala2016,Sorce2016}, the cosmic density from Lyman-$\alpha$ tomographic observations \citeapp{Horowitz:2019}, and also more recently to study massive galaxies \citeapp{2020MNRAS.493.4607R}

In this work, we extend this concept to Cosmic Noon by running $N$-body simulations based on the COSMOS initial conditions introduced in section for initial conditions  using the massively parallelized PKDGRAV3 $N$-body code \citeapp{Potter2017}, which utilizes the Fast Multipole Method for gravity calculations and has been tested in a 2 trillion particle cosmological simulation.
MUSIC returns the initial conditions files in the TIPSY format, that can be directly read in PKDGRAV3. 
In this study, we are mainly interested in the galaxy cluster evolution from the observed redshifts of the galaxy surveys, at $z = 2.3$, until $z=0$.

However, to precisely describe the observed structures over the time-evolution of their lightcones, we also output six PKDGRAV3 snapshots in the interval of $z=2-2.5$ with a separation of $\Delta z = 0.1$.

\begin{figure}[tb!]
\centering 
\includegraphics[width=.49\textwidth]{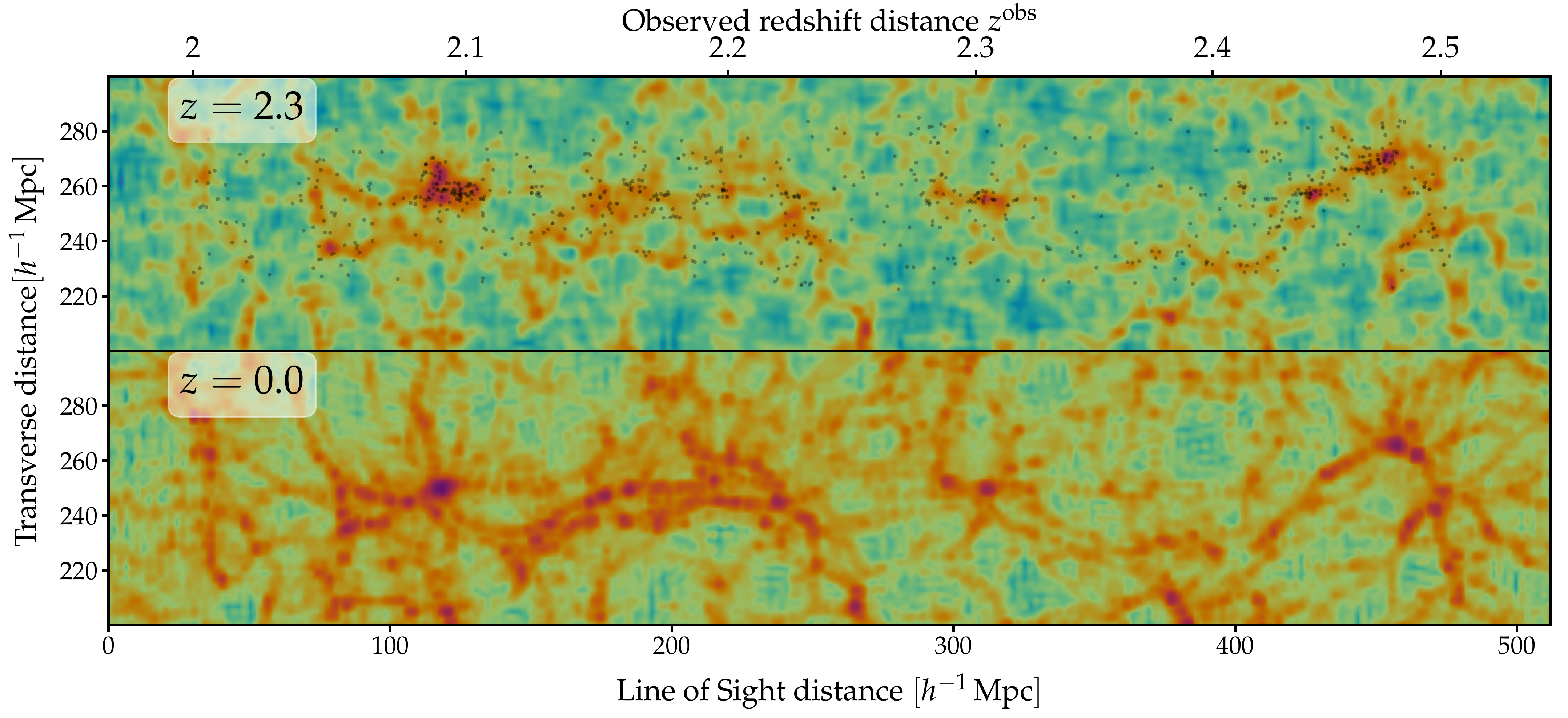}
\includegraphics[width=.49\textwidth]{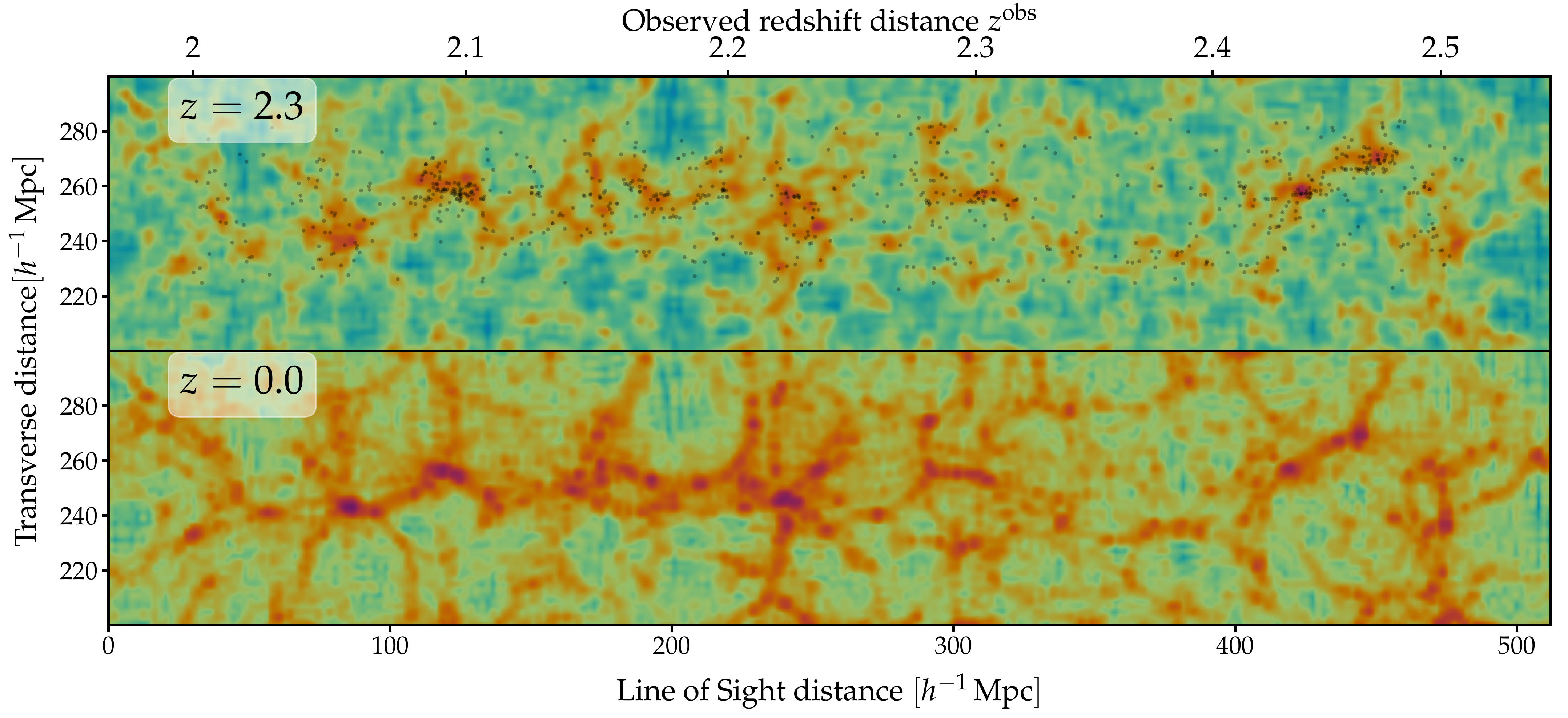}
\includegraphics[width=.49\textwidth]{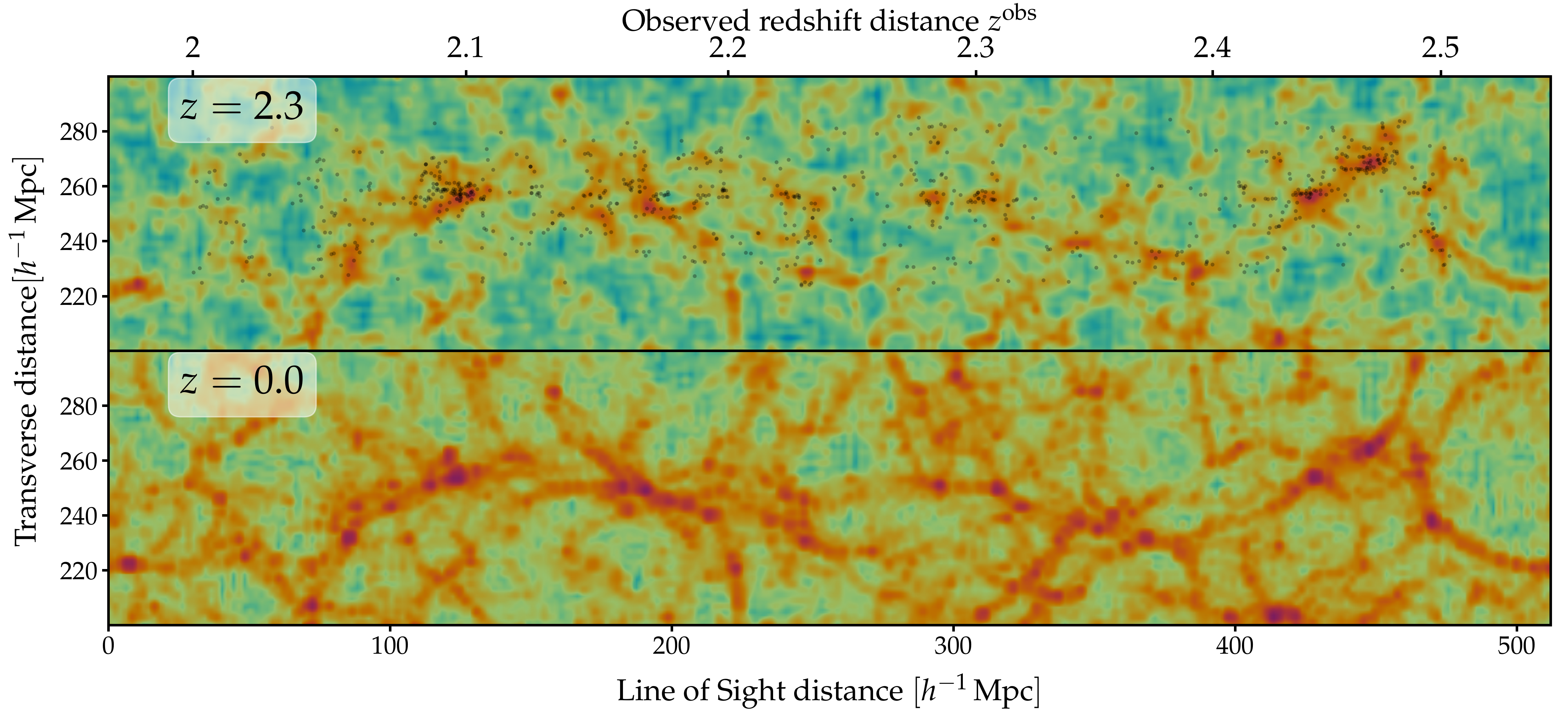}
\includegraphics[width=.49\textwidth]{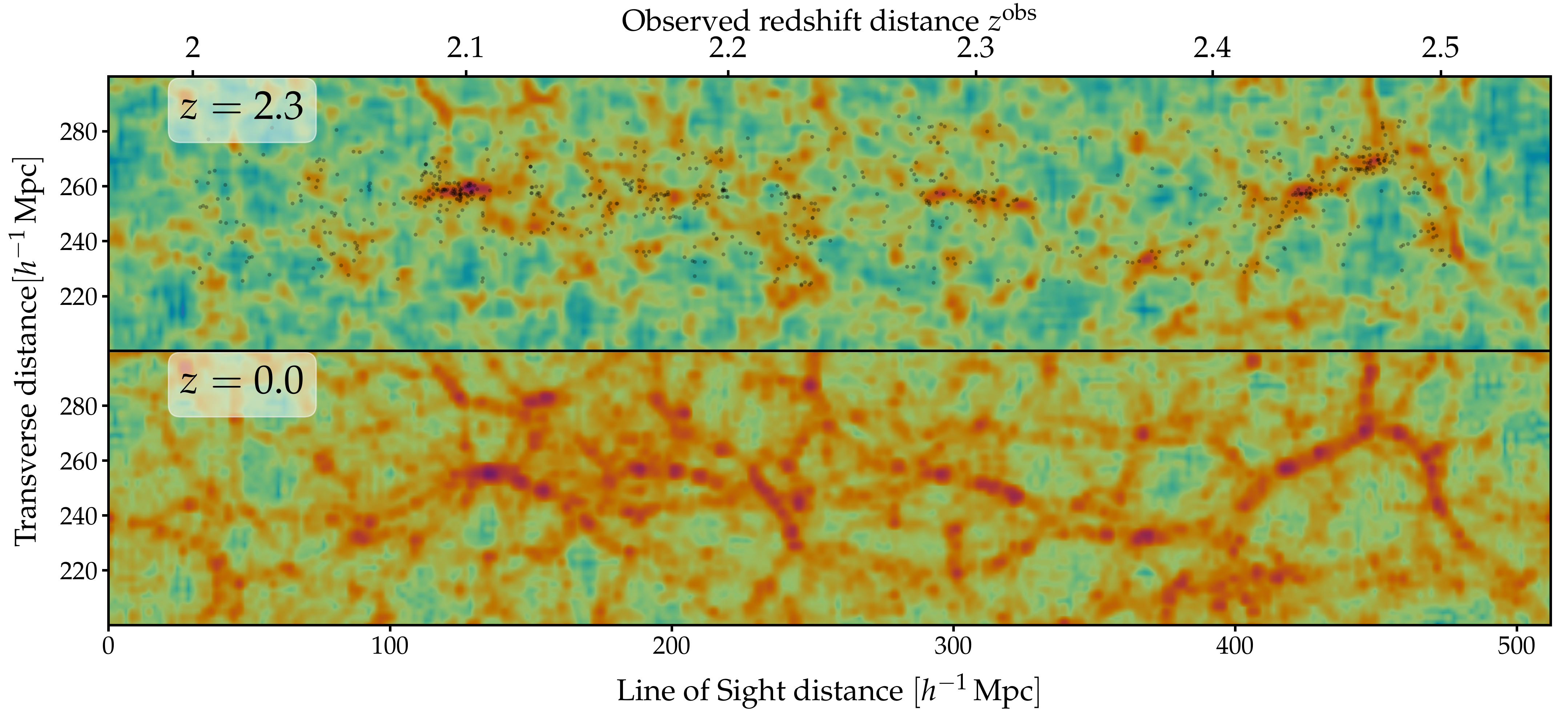}
\caption{\label{fig:snapshots} Pairs of matter density field slice plots showing the redshift snapshot of $z=2.3$ (which is the mean redshift of the input galaxies) on the top panel and the $z=0$ snapshot on the bottom. We signpost the galaxy positions with black dots on top of the $z=2.3$ snapshot.  We transformed the galaxy surveys to a coordinate frame where the line of sight direction is aligned with the abscissa. }
\end{figure}

In Supplementary Figure \ref{fig:snapshots} we show four different constrained simulations in pairs of $z=2.3$ and $z=0$.   We also display the galaxy positions with black dots on top of the $z=2.3$ snapshot.
The density contrast of the $z=2.3$ snapshot closely follows the galaxy distribution in all cases. Outside the observed region, i.e. $\sim\pm 20\hmpc$ from the edges along line of sight and $\sim\pm 10\hmpc$ in transverse direction, the realizations have been substituted by a random field created from the $\Lambda$CDM power spectrum.
Differences in the small-scale structure of the realizations, in addition to the variation of random fields outside the observed area, lead to notably different final density fields at $z=0$.
However, we see consistently forming large-scale structure, such as the strong overdensity centered at $z^{\rm obs} \sim2.1$, and an extended filamentary structure centered at $\langle z^{\rm obs}\rangle \sim2.45$. Other, less well constrained
regions in the observed volume, show that the $z=0$ outcome can be very different. For example, within the four $\zav$ snapshot shown in Supplementary Figure~\ref{fig:snapshots}, we can see different fates of one of the overdensities at $\zobs \sim2.05$: in one case, it collapses into a singular cluster at $z=0$; in two cases becomes part of a wider overdense region; and in one case diffuses into a shallow density. Also, we can see that in the region between $2.12 \lesssim \zobs \lesssim 2.18$, a few density peaks consistently form at the $\zav$ snapshots. However, their final state can be a singular high density filament at $z=0$, or, several aligned filaments that run in parallel to each other. Finally, it is worth mentioning that even though in all realizations the \textit{Hyperion} supercluster is forming a massive filament, the internal structure changes. We can see a different number of member halos forming within the filament, mostly consisting of 4-5 massive halos (see also Figure \ref{fig:hyperion} for a better visual impression). 
Future deep surveys with higher galaxy sampling compared to our dataset, in conjunction with improved modelling of redshift-space distortions and other effects, should allow more consistent predictions on the evolution of these structures.

We also tested for a possible systematic influence of the simulations' grid resolution on the final results. 
To do so, we increase the resolution of our initial conditions from $N_{\rm C}= 256^3$ to $1024^3$ by augmenting small-scale modes in the Gaussian random field, while preserving the large-scale ones. This procedure is justified since we are interested in the most massive halos of these simulations, regardless of the halo substructures.
The results of the high-resolution simulation is consistent with those of the low-resolution run performed
with the same initial conditions, yielding clusters that agree to within $\sim2\%$ in position and virial mass. This is clearly a subdominant effect as compared to the variance among the different realizations, and we consider the grid resolution of $N_{\rm C}= 256^3$ to be converged for our purposes. The resemblance is shown in Supplementary Figure \ref{fig:res}.
\begin{figure}[tb!]
\centering 
 \includegraphics[width=.73\textwidth]{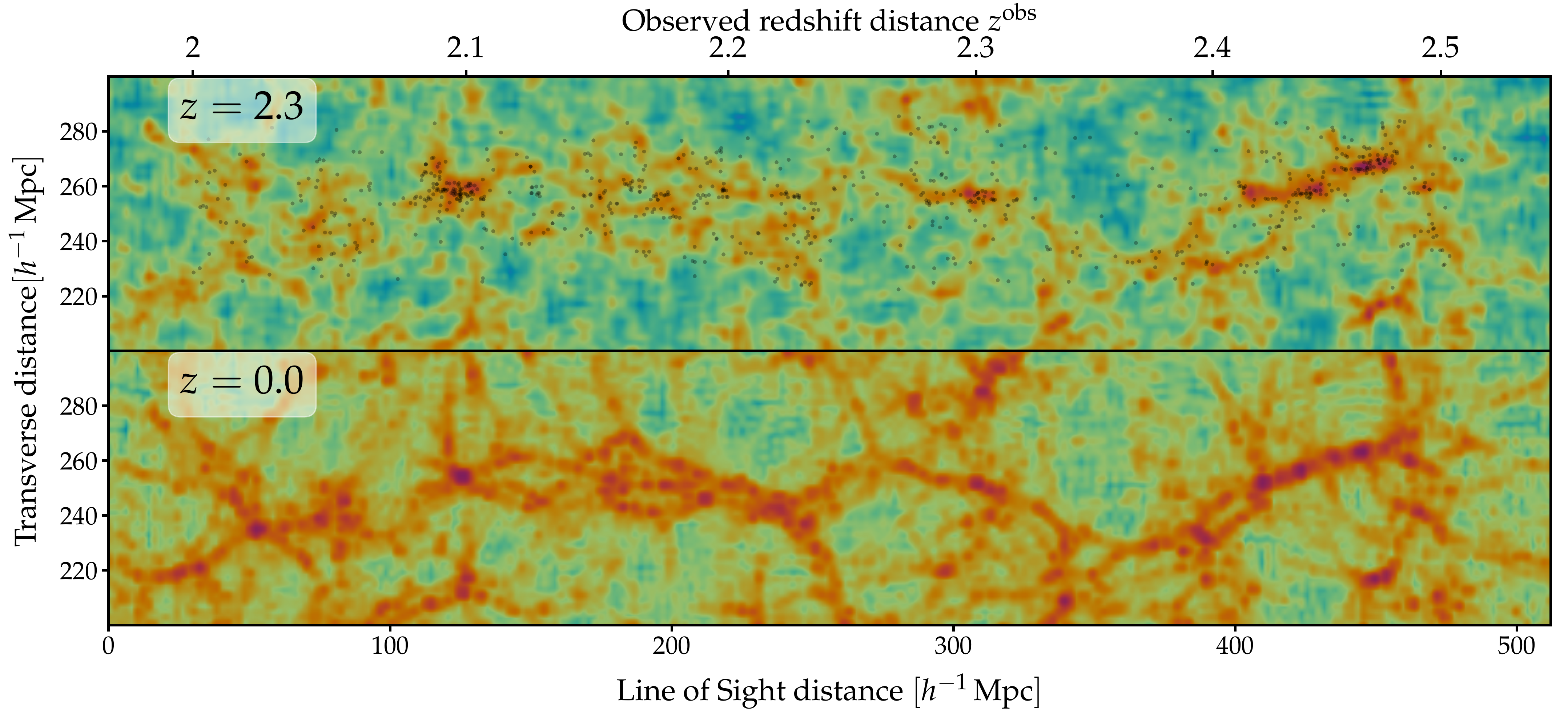}
  \includegraphics[width=.73\textwidth]{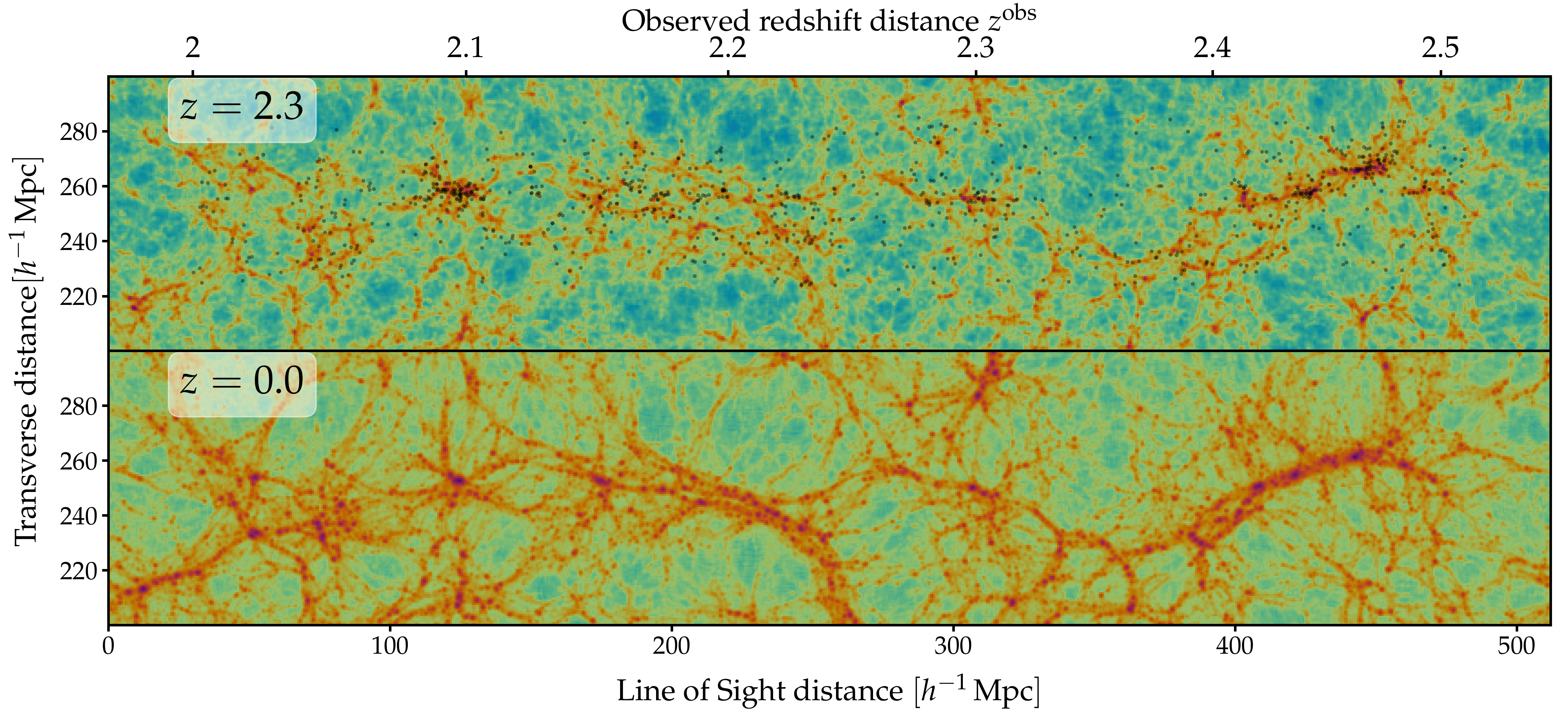}
\caption{\label{fig:res} Comparison of the $256^3$ (top) and the $1024^3$ (bottom) simulation for the same initial conditions. We find a clear agreement of the large-scale structure of both simulations and a matching of cluster masses and positions with a $\sim2\%$ accuracy.}
\end{figure}

\subsection*{Halo Analysis}
\label{sec:halos}

We identify halos from the PKDGRAV3 simulation snapshots using the ROCKSTAR phase-space halo finder  \citeapp{rockstar}. Subsequently, we calculate the merger trees across the PKDGRAV3 snapshots using the consistent-trees package \citeapp{consistenttrees}.
We then select halos at $z=0$ that have a minimum virial mass of $M_{\rm vir} \geq 2\times 10^{14}\hmsun$ \citeapp{Chiang2017}.
The theoretical $\Lambda$CDM halo mass function at $z=0$ predicts $N_{\rm Halos} \sim 26$ halos of $M_{\rm vir} \geq 2\times 10^{14}\hmsun$ in the considered COSMOS field volume \citeapp{hmf} (see also Supplementary Figure \ref{fig:hmf}).

Next, we trace the $z=0$ dark matter particles that ROCKSTAR associates with each halo back to $z=2.3$, allowing us to self-consistently define a protocluster as the set of all progenitor dark matter particles of a $z=0$ halo in its Lagrangian volume.
The center-of-mass positions for each protocluster (in the observed volume) are shown with blue diamonds in the upper panel of Figure \ref{fig:dens}, while two examples of the virialized halos and their Lagrangian volumes are shown in Figure \ref{fig:halos}.

\subsubsection*{Matching the LRPCs}

Protocluster searches are a very important research topic for understanding structure formation at high redshifts \citeapp{protoclustersearch1,protoclustersearch2,protoclustersearch3,protoclustersearch4,protoclustersearch5,protoclustersearch6}.   

In this section, we  match the protocluster's center-of-mass position to the LRPCs for all realizations by performing a nearest-neighbor search.
For the choice of the search radius we must take into account the following conditions:
\begin{itemize}
	\item \textbf{Cosmological displacements:} 
	\\ The most massive halos, that grow in an isolated environment, grow nearly in-place with a final position at $z=0$ that is nearly identical to the center-of-mass position of the protocluster.
    In contrast, less massive halos in dense environments show significant displacements from $z=2.3$ to $z=0$ that can be of order $\sim10 \hmpc$.
	\item  \textbf{Incomplete information:} \\ The observed galaxy catalogs contain limited information in two aspects. First, the phase-space distribution of the galaxies is incomplete, as we only have spatial information. The unknown peculiar velocities result in a distortion along line of sight, called redshift space distortions (RSD). Also, the velocity autocorrelation of galaxies at cosmological distances of $r=100 \hmpc$ can still be high \citeapp{Turner2021}. Consequently, structures outside of our observed volume have a significant influence on the velocities of our halos, called super-survey modes in the literature  \citeapp{Takada2013}.
	\item  \textbf{Sampling variance:} \\ The different initial conditions inferred with the COSMIC BIRTH code are individual Markov-Chain realizations drawn from a posterior probability distribution. This chain variance depends on the statistical model, as well as on the marginalized nuisance parameters \citeapp{Kitaura2019, Ata2021, Ata2015}. Therefore, we naturally expect a certain variance within the Markov-Chain samples. \\
\end{itemize}
To account for these displacements we define a search radius of $r_{\rm search}= 15\hmpc$ around the LRPCs and scan for the center-of-mass positions of the $z=2.3$ protoclusters in their vicinity. Details on the robustness and the justification for this search radius are given below.
If multiple candidates are found, we choose the most massive among them, or, if their Lagrangian volume overlaps at $z=2.3$, we sum their masses at $z=0$.
If a matching of protoclusters and LRPCs is successful, we show the distance as white line in the upper panel of Figure \ref{fig:dens}.

\subsubsection*{COSTCO Protoclusters}

Besides identifying the LRPCs that were previously known, we look for other recurringly forming halos within the COSTCO suite.
For this purpose, we stack all halos of the COSTCO multiverse into one volume and apply the \texttt{Density-Based Spatial Clustering of Applications with Noise} (DBSCAN) algorithm, implemented in the  \texttt{Scikit-learn} package.
This algorithm identifies clustered points given the halo density of the COSTCO multiverse and allows to model noise of the point cloud with an additional parameter. 
To tune this noise parameter, we first apply DBSCAN to the LRPCs only. Once we find a best fit setup such that the LRPCs are robustly identified, we apply the exact same scheme to the remaining volume and look for additional clusters in the COSTCO multiverse.
We then use the center-of-mass position of these newly identified clusters as anchor points and scan the 50 realizations at these positions for massive halos.
This has also been studied with the MIP suite using a signal-to-noise ratio \citeapp{Aragon-Calvo2016}. MIP is a set of correlated cosmological simulations and is based on random initial conditions. The initial conditions of the MIP realizations are consistent in their large-scale Fourier modes, but were randomized on the small scales. This represents an interesting analogy to the constrained COSTCO simulations.
We show the corresponding confidence map of the identified protoclusters on the top panel of Supplementary Figure \ref{fig:heatmap} for all 50 realizations stacked into the same volume. We write the names of the LRPCs and COSTCO protoclusters (the same order as in Table \ref{tab:proto_obs}) on top of the identified regions.
It is clearly noticeable that the ZFIRE and Hyperion structures show the strongest significance, while the other structures have more extended and less distinct contour regions, which means that the uncertainty of these structures is higher. Nevertheless, all identified COSTCO protoclusters form in the majority of our realizations. This shows that constrained simulations are a well-suited method for the identification of less massive protoclusters, that would be less efficient with other methods.

The bottom panel of Figure  \ref{fig:dens} shows the $z=0$ snapshot and the two kinds of encircled halos. White circles indicate halos for which a protocluster-LRPC matching was successful. This means that the white circles represent the predicted $z=0$ state of the LRPCs.
Blue circles show massive halos that could not be linked to a LRPC partner and are newly discovered by our analysis. 

On the bottom panel of Figure \ref{fig:heatmap} we show the mean  density of all 50 COSTCO $z=2.3$ snapshots. While densities outside the observed volume average out, we see an excellent agreement of the averaged structures and galaxy distribution within the constrained volume.
Interestingly, on average COSTCO~V arises from a stronger overdensity than COSTCO~II, but does not evolve to a higher mass than COSTCO~II (Table~\ref{tab:proto_obs}). The main difference  of these two structures lies in their respective large-scale environments: COSTCO~II forms in a more isolated environment and can accrete surrounding matter without disruption, whereas COSTCO~V is within the vicinity of the Hyperion filament/supercluster and likely experiences some form of large-scale mass stripping by its more massive neighbor. This stresses the importance of the large-scale environment for determining the evolution of a protocluster rather than purely relying on local galaxy number density.

\subsubsection*{Robustness of the Halo Identifications}
We asses the robustness of our protocluster identifications. In particular, we want to know the probability of finding a massive halos at any given fixed position in a search radius of $r_{\rm Search} = 15 \hmpc$ within our simulations by chance.
The theoretical $\Lambda$CDM halo mass function predicts about  $\mathrm{n}_{\rm H}~\sim 8\times 10^{-6}\,h^{3}\,\rm{Mpc}^{-3}$ halos per unit volume with $M_{\rm vir} \geq 2\times 10^{14}\hmsun$.
Thus, we estimate the expected number of these halos in a volume of 
$V_{\rm H}= 4/3 \,\pi \,r_{\rm Search}^3 $ to be
$V_{\rm H} \times \mathrm{n}_{\rm H} \sim  0.1 $.
Thus, assuming that the halo distribution in 50 separate volumes is an independent and identically distributed random variable, we expect 5 halo detections in this search radius by chance.

To confirm this estimate numerically, we ran five simulations with the same resolution and same set of cosmological parameters as COSTCO, but with random initial conditions. We then chose 10 fixed positions  inside these random simulations and 
searched for $M_{\rm vir} \geq 2\times 10^{14}\hmsun$ halos in a radius of $r_{\rm Search} = 15 \hmpc$, ensuring that these points were separated by at least $100\hmpc$. On average we found $\sim 0.5$ halos per position. Extrapolating this number to 50 realizations, we therefore expect to find $\sim 5$ halos by chance at a random position inside the COSTCO suite. This agrees with the analytical estimate from the halo mass function.

Relying on Poisson statistics, we consider a COSTCO halo to be a robust $5\,\sigma$ detection, if its counterparts are found in at least $\geq 27.5$ realizations at the same position given the search radius of $15\hmpc$.

All our claimed COSTCO protoclusters shown in Table \ref{tab:proto_obs} fulfill this
detection threshold, except for COSTCO J095945.1+020528 which was found in 27 out of 50 realizations ($4.9\,\sigma$), but we have nevertheless decided
to include it as a detection. Note that apart from this object and the other listed protoclusters, 
the next most significant candidate is found in only 8 out of 50 realizations ($1.45\,\sigma$).
This further validates our decision to include COSTCO J095945.1+020528, since it clearly rises above the random background noise.

For the robust identification of halos with smaller masses, we would need to shrink the search radius in order not to be contaminated by random detections. However, this requires a better knowledge of the $z=2.3$ to $z=0$ displacement field, which in turn requires a higher number density of tracers that constrains the initial conditions. 
In a follow-up paper, we will forecast the sensitivity of our methods toward lower-mass clusters in future high-redshift spectroscopic surveys.

\subsubsection*{COSTCO Halo Mass Function}

We compare the halo mass functions (HMF) from the COSTCO realizations to the five random simulations and also to the theoretical HMF that we calculated with the \texttt{HMFcalc} package. As we can see in Supplementary Figure \ref{fig:hmf}, the COSTCO as well as the random initial conditions simulations show close agreement with the theoretical HMF within the uncertainties. Also we can see that the high mass end at $M_{\rm{Halo}} > 10^{15}\,\hmsun$ gets very noisy due to the low number of halos. 
\begin{figure}[tb!]
\centering 
 \includegraphics[width=.9\textwidth]{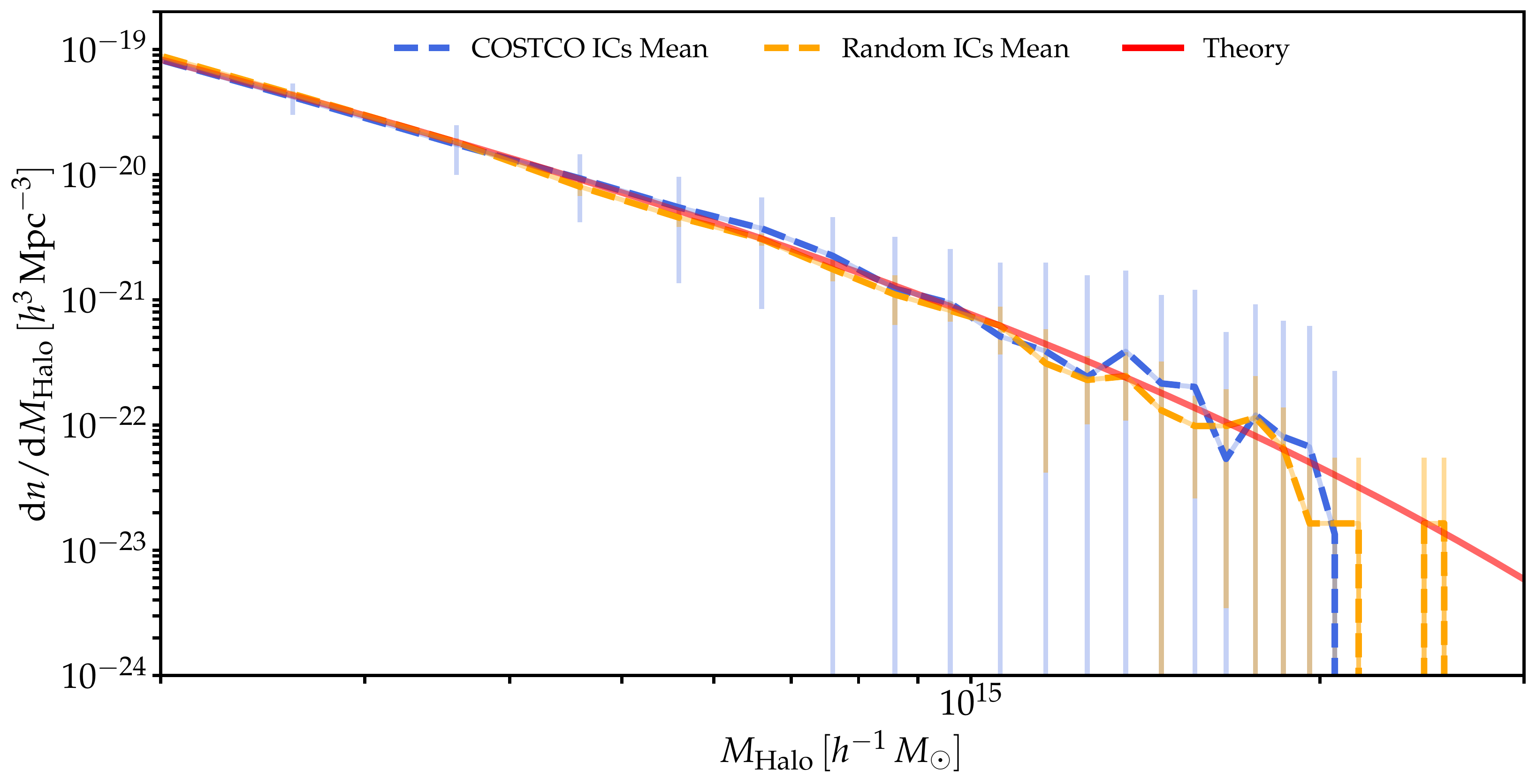}
\caption{\label{fig:hmf} Mean halo mass function of the COSTCO runs (blue dashed line), five simulations with random initial conditions (orange dashed line) and the theoretical prediction (red line). Error bars in the same color as solid lines.}
\end{figure}

Within the statistical uncertainty of the HMF, we can see that the COSTCO simulations suite is consistent with the $\Lambda$CDM predictions. However, in contrast to the random simulations, the COSTCO HMF shows two bumps in the high-mass tail that slightly exceed the random simulations. We find that these bumps are caused by the ZFIRE and Hyperion cluster and may hint towards an overabundance of massive halos in the studied volume. We will the study the significance of the massive halos in more detail in a subsequent work.

\subsubsection*{Effective Scale of the COSTCO Realizations}

The COSTCO simulation ensemble shares the same large-scale structure, while small scales can notably be different. 
Therefore, we evaluate the effective scale on which the different realizations are correlated. 

This is done as follows.
\begin{itemize}
    \item We compute the mean of all 50 COSTCO $z=2.3$ snapshots (see Supplementary Figure \ref{fig:heatmap} bottom panel). We select a rectangular volume that contains the observed region.
    \item We apply Gaussian smoothing with kernel sizes of $r_{\rm S} = 3-6 \hmpc $ on a simulation with random initial conditions and select the same volume.
    \item We compute the power spectra of the COSTCO mean field and the variously smoothed random simulation in the selected volume.\\
\end{itemize}
\begin{figure}[tb!]
\centering 
 \includegraphics[width=.9\textwidth]{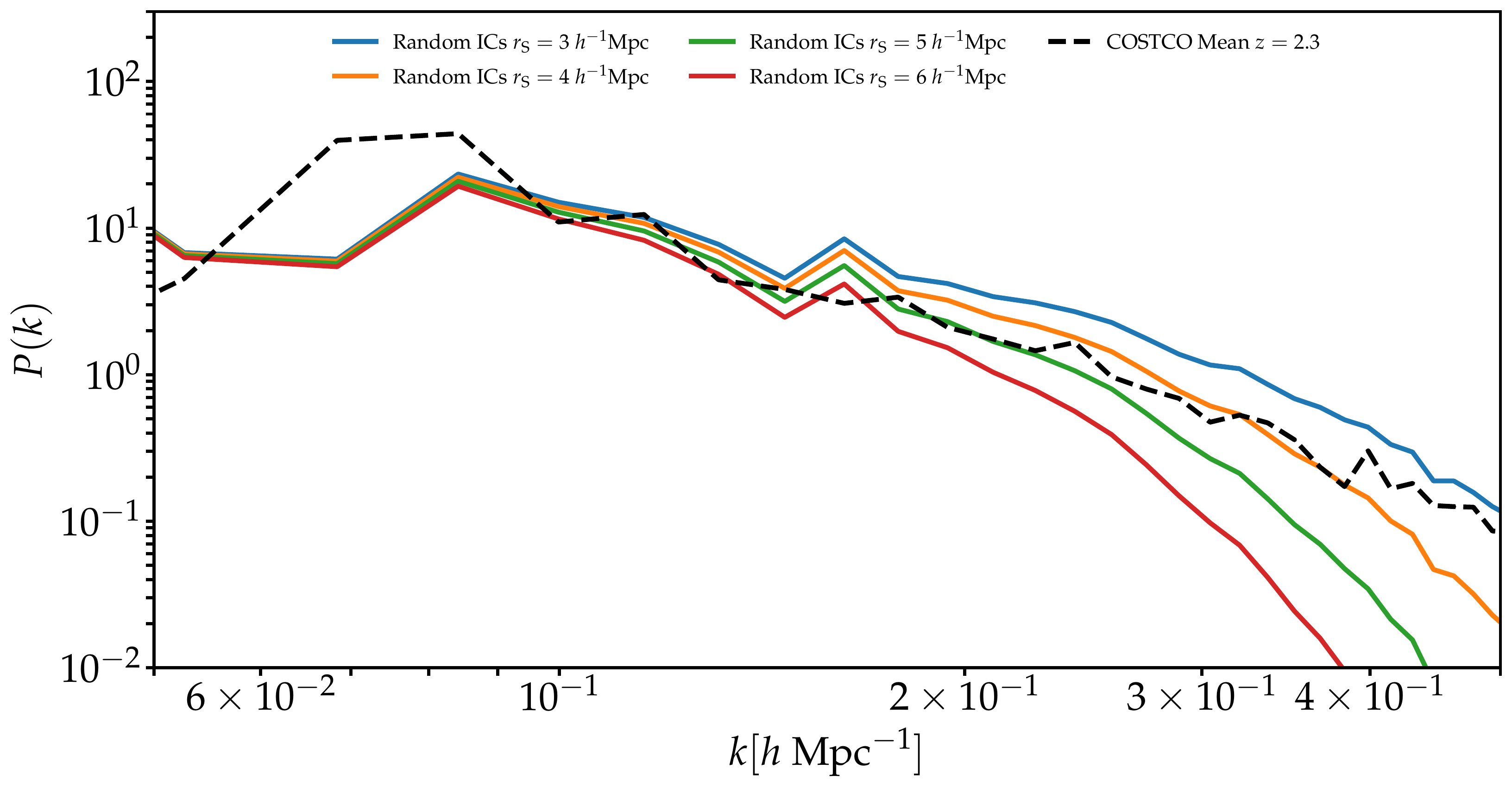}

\caption{\label{fig:mean} Assessment of the effective scale calculated from the COSTCO simulations with the power spectra of the COSTCO mean field (dashed black line) and the smoothed reference simulation with random initial conditions (solid colored lines). The COSTCO realizations agree on scales that correspond to a Gaussian smoothing of $r_{\rm S} \sim 4.5 \hmpc$.}
\end{figure}

We show the resulting power spectra in Supplementary Figure \ref{fig:mean}.
The dashed black line shows the power spectrum for the COSTCO mean field, while the solid lines show the different smoothing scales of the reference simulation that we ran with random initial conditions. 
Comparing the power spectra  we can directly read off, that the COSTCO simulations are effectively correlated on scales of $r_{\rm{S}} \sim 4.5 \hmpc$ (compare green and orange solid line) while fluctuating for smaller scales. The correlated scales are well suited to analyze the formation and fate of galaxy protoclusters, focusing on the final masses and positions.

\section*{Software}
This research made use of matplotlib \citeapp{Hunter:2007}, SciPy \citeapp{Virtanen_2020},  Scikit-learn \citeapp{scikit-learn}, Astropy \citeapp{2018AJ....156..123A, 2013A&A...558A..33A}, HMFcalc \citeapp{hmf}, Dark Emulator \citeapp{Nishimichi2019}, pynbody \citeapp{pynbody}, and ytree \citeapp{ytree}.

\bibliographystyleapp{sn-mathphys}
\bibliographyapp{sn-bibliography}
\small
\section*{Acknowledgements}
The authors thank Peter Behroozi and Doug Potter for their support with Rockstar and PKDGRAV3, respectively. MA thanks Takahiro Nishimichi, Masahiro Takada, and Valeri Vardanyan for useful discussions.
MA was supported by JSPS Kakenhi Grant JP21K13911.
KGL acknowledges support from JSPS Kakenhi Grants JP18H05868 and JP19K14755.
CDV acknowledges support from the Spanish Ministry of Science and
Innovation (MICIU/FEDER) through research grants PGC2018-094975-C22 and
RYC-2015-18078.

This research is based on observations undertaken at the European Southern Observatory (ESO) Very Large Telescope (VLT) under Large Program 175.A-0839 and has been supported by the Swiss National Science Foundation (SNF).
Some of the material presented in this paper is based upon work supported by the National Science Foundation under Grant No. 1908422.
This work was made possible by the World
Premier International Research Center Initiative (WPI),
MEXT, Japan.

\section*{Author contributions statement}
MA calculated the selection functions, initial conditions, simulations, halo catalogs, and conducted the final analysis. MA and KGL conceptualized the analysis and wrote the paper. CDV helped performing the simulations. FSK provided expertise about density field reconstructions. OC, BCL, and DK provided the (partly) proprietary data and relevant literature. TM provided the visualizations.
All authors reviewed the manuscript. 

\section*{Data availability}
The data products generated for this study are available at \url{https://zenodo.org/record/6425013}.

\section*{Software availability}
Analysis codes used for this study are available at \url{https://github.com/gmetin/COSTCO}.

\section*{Competing Interests} 
The authors declare that they have no competing financial interests.

\section*{Correspondence} 
Correspondence and requests for materials can be addressed to the corresponding author M.A. (email: metin.ata@ipmu.jp)

\end{document}